\documentclass[aps.pra,twocolumn,showpacs,preprintnumbers,amsmath,amssymb]{revtex4}

\usepackage[utf8]{inputenc}
\usepackage[english]{babel}

\usepackage{setspace}
\usepackage{booktabs}
\usepackage{multirow}
\usepackage{amsmath,amssymb}
\usepackage{qcircuit}
\usepackage{caption}
\usepackage{lipsum}

\usepackage{natbib}
\usepackage{stfloats}
\usepackage{graphicx}
\usepackage{subcaption} 

\usepackage{physics}
\usepackage[figurename=Fig., justification=RaggedRight]{caption}
\usepackage{amsmath}
\usepackage{xcolor}
\usepackage{hyperref}

\begin{document} 

\title{Dynamical quantum Ansatz tree approach for the heat equation}

\author{N. M. Guseynov$^{1,2}$, W. V. Pogosov$^{1,2,3}$, A. V. Lebedev$^{1,2}$}
\affiliation{$^1$Dukhov Research Institute of Automatics (VNIIA), Moscow, 127030, Russia}
\affiliation{$^2$Advanced Mesoscience and Nanotechnology Centre, Moscow Institute of Physics and Technology (MIPT), Dolgoprudny, 141700, Russia}
\affiliation{$^3$Institute for Theoretical and Applied Electrodynamics, Russian Academy of Sciences, Moscow, 125412, Russia}

\begin{abstract}

Quantum computers can be used for the solution of various problems of mathematical physics. In the present paper, we consider a discretized version of the heat equation and address its solution on quantum computer using variational Anzats tree approach (ATA). We extend this method originally proposed for the system of linear equations to tackle full time dependent heat equation. The key ingredients of our method are (i) special probabilistic quantum circuit in order to add heat sources to temperature
distribution, (ii)  limiting auxiliary register in the preparation of quantum state, (iii) utilizing a robust cluster of repetitive nodes in the anzats tree structure. We suggest that our procedure provides an exponential speedup compared to the classical algorithms in the case of time dependent heat equation. 

\end{abstract}

\maketitle

\section{INTRODUCTION}\label{s:intro}

The idea of a quantum computer has been originally proposed in the context of the simulation of quantum systems, which cannot be efficiently studied using classical computation methods due to the exponential size of the Hilbert space \cite{FEYNMAN}. After that, the number of possible applications of quantum computers has increased drastically. Among known quantum algorithms, one can mention Shor's factorization algorithm \cite{math_apply,Shor} and Grover's search algorithm \cite{grover,q_search}, which provide computational speedup compared to its classical analogs. In addition, during last years, certain  progress has been achieved in the development of quantum computing algorithms for the solution of mathematical equations \cite{jin2023quantum,herman2023quantum,baiardi2023quantum,hassija2020forthcoming}. One of the examples is the heat equation \cite{guseynov2023depth,linden2022quantum,pollachini2021hybrid}, which is of importance for many real-world applications \cite{perona1990scale,carslaw1959conduction,wilmott1995mathematics}. 

A closely related task is the solution of  linear algebra problems, such as systems of linear equations, for which quantum computers can be also used. The most well-known example is the Harrow, Hassidim, and Lloyd (HHL) quantum algorithm \cite{PhysRevLett.103.150502,duan2020survey,wang2020quantum,montanaro2016quantum}, which provides exponential speedup for solving sparse linear systems\cite{clader2013preconditioned}. Alternatively, the variational approaches\cite{bravo2019variational,cerezo2021variational} can be advantageous in solution of the same or similar tasks especially in the era of noisy intermediate scale quantum (NISQ) devices \cite{cerezo2021variational,fontanela2021quantum,fontana2021evaluating,kubo2021variational,lubasch2020variational,yang2021variational} characterized by noticeable quantum noise rates as well as by other drawbacks \cite{Preskill2018quantumcomputingin,wang2021noise,holmes2022connecting}. As expected, these difficulties can be resolved with the fault tolerant quantum computation. However, this requires a large overhead in the qubit number due to the necessity to encode a single logical qubit into many physical qubits with  sufficiently low noise rates\cite{devitt2013quantum,fowler2012surface}. There are several physical realizations of quantum computers and the most successful are the ion \cite{pino2020demonstration,grzesiak2020efficient} and superconducting\cite{bravyi2022future,gambetta2020ibm} platforms. However, in general, state-of-the-art technologies do not allow yet to implement fault tolerant quantum computation. It has very recently, however, been argued and demonstrated using superconducting quantum processor that the utility of quantum computation is possible without full fault tolerance, i.e., within the NISQ paradigm, using quantum error mitigation technique applied to the digital quantum simulation task \cite{kim2023evidence}.

We have recently analyzed \cite{guseynov2023depth} several quantum variational algorithms for solution of the heat equation. We have restricted ourselves to the case of a single time step, for which the problem is reduced to the solution of a linear system  with a tridiagonal matrix. It turns out that the most promising algorithm is the one we addressed as the ATA (Anzats Tree Approach) originally proposed earlier in Ref. \cite{huang2021near} which  provides an exponential speedup compared to classical state-of-the-art algorithms.   The general idea of the ATA is to utilize a hierarchical optimization technique where the form of the anzats explicitly depends on the form of the linear system under study. 

The aim of the present paper is to address a full time evolution of the time-dependent heat equation using the ATA and develop an efficient scheme for the solution of this equation on a quantum computer. We find that a straightforward application of the ATA to the time dependent heat equation requires an addition of heat sources to a temperature distribution, which turns out to be a highly nontrivial task in the quantum computing paradigm. In order to solve this problem we make use the ATA and construct the Anzats tree which generates the spatial heat source distribution as a quantum state encoded into a quantum register.

In addition, we study the hierarchical structure of the ATA and find that there exists a robust cluster of repetitive nodes in the Anzats tree solution. Remarkably, this feature can be utilized to reduce the computational cost. 

Finally, we compare our method with HHL algorithm performance. We find that HHL algorithm suffers from the exponential decay of the sequential success probability, as the time grows. Moreover, HHL requires longer quantum circuits compared to our approach. On the other hand, our variational approach reveal the opportunity of achieving exponential speedup for solution of the time dependent heat equation.  

The paper is organized as follows. In Section \ref{s:prelim}, we present the heat equation as well as its discretized form, and then we introduce ATA and its realization for a single time step. In Section \ref{s:accuracy for time dependent}, we apply ATA in a straightforward way to the time dependent heat equation and discuss an optimal choice of control parameters. In Section \ref{s:repetitive}, we analyze the structure of the ansatz tree and propose a method to reduce the computational cost. In Section \ref{ch:dropout}, we describe a tree reduction procedure that allows us to achieve quantum speedup. Section \ref{s:HHL} presents a comparison with HHL algorithm. We conclude in Section \ref{s:conclusion}. 

\section{Preliminaries}\label{s:prelim}
\subsection{Heat equation}

In this paper we consider the heat equation with constant coefficients, a given initial temperature distribution, and periodic spatial boundary condition,
\begin{eqnarray}
\begin{gathered}
a^2 \Delta U(\vec{r},t)-\frac{\partial U(\vec{r},t)}{\partial t}=f(\vec{r},t);\\
U(\vec{r},0)=\chi(\vec{r}); \qquad U(\vec{r},t)=U(\vec{r}+\vec{R},t),
\end{gathered}
\label{Thermal_conductivity_equation}
\end{eqnarray}
where $f(\vec{r},t)$ and $\chi({\vec{r}})$ are the heat sources and initial temperature distribution, respectively. Hereafter we demonstrate the application of our algorithm to the one-dimensional heat equation
\begin{eqnarray}
\begin{gathered}
a^2\frac{\partial^2 U(z,t)}{\partial z^2}-\frac{\partial U(z,t)}{\partial t}=f(z,t),\\
U(z,0)=\chi(z),\qquad U(z,t)=U(z+Z,t).
\label{1_dimensional_thermal_conductivity_equation}
\end{gathered}
\end{eqnarray}

We argue that the one-dimensional case is fundamental, and the extension to the multidimensional case is straightforward and described in Appendix \ref{app:high dim}. We notice that the multidimensional case requires linearly more qubits, which makes it difficult to perform the corresponding simulation using a classical computer.

Next, we use an implicit finite-difference scheme \cite{zienkiewicz2005finite,ozicsik2017finite}, which is a classical tool for the numerical solution of differential equations
\begin{eqnarray}
    &&(-2-c)U^{\tau+1}_i+U^{\tau+1}_{i+1}+U_{i-1}^{\tau+1}=b_i^\tau
    \label{equation_grid}
    \\
    &&\qquad\qquad b^\tau_i= c\bigl( \delta t\, f_i^\tau-U_i^\tau\bigr),
    \nonumber
\end{eqnarray}
where the lower index $i=0,\dots,N_z$ refers to the spatial grid and the upper index $\tau = 0,\dots,N_t$ denotes the temporal grid, $U_i^0 = \chi_i$ and $U_0^\tau = U_{N_z}^\tau$, $c=(\delta z)^2/(a^2 \delta t)$ with $\delta z$ and $\delta t$ are the spatial and temporal grids resolutions respectively.  The usage of the implicit scheme yields the stability of the solution for arbitrary parameters of the equation and the grid size \cite{higham2002accuracy}. Thus, the transition from the consecutive time instants $U_i^\tau \to U_i^{\tau+1}$ is done by solving the linear system,
\begin{eqnarray}
Ax=b,
\label{Ax=b}
\end{eqnarray}
where
\begin{eqnarray}
A(c)=\left(\begin{array}{ccccc}
-2-c & 1  & \dotsm & 0 & 1 \\
1 & -2-c  &\dotsm& 0 & 0\\
\rotatebox[origin=c]{270}{\dots}&&\rotatebox[origin=c]{-45}{\dots}&&\rotatebox[origin=c]{270}{\dots}\\ 
0 & 0 &\dotsm & -2-c &1\\
1 & 0 &\dotsm & 1 &-2-c\\
\end{array}
\right),
\label{A_definition}
\end{eqnarray}
\begin{eqnarray}
x=\left(\begin{array}{c}
U_0^{\tau+1}  \\
U_1^{\tau+1} \\
\rotatebox[origin=c]{270}{\dots}\\
U_{N-1}^{\tau+1}\\
\end{array}
\right),\qquad b=\left(\begin{array}{c}
b_0^{\tau}  \\
b_1^{\tau} \\
\rotatebox[origin=c]{270}{\dots}\\
b_{N-1}^{\tau}\\
\end{array}
\right).
\label{Axb}
\end{eqnarray}

\subsection{The ATA solution of a linear system}\label{sub:linear system}

The Anzats tree approach solution of a general linear system $Ax=b$ with a hermitian matrix $A$ exploits the unitary decomposition of the matrix $A$,
\begin{eqnarray}
A=\sum_{i=1}^{K_A}\beta_iU_i,
\label{A_ansatz_tree_decomposition}
\end{eqnarray}
where $U_i$ are unitary operators. Such decomposition is always possible as far as any hermitian matrix of a size $2^n \times 2^n$ can be decomposed into a sum of Pauli products: $U_i = \hat\sigma_{\alpha_1} \otimes \cdots \otimes \hat\sigma_{\alpha_n}$, $\hat\sigma_{\alpha_j} \in\{I,X,Y,Z\}$.  However, the Pauli products decomposition is not efficient since it requires $K_A \sim 2^n$ Pauli strings in general.  The efficient decomposition implies that: (i) the unitary operators of the decomposition can be presented by quantum circuits with a polylogarithmical depth $\propto n^p$, (ii) the number of $U_i$ also scales polylogarithmically with the size of the matrix $A$. 

In the ATA one minimizes the loss function, which is the well-known $\ell_2$-norm loss used in regression methods,
\begin{eqnarray}
L_R(x)=\| Ax-\ket{b}\|^2_2= x^\dagger A^\dagger A x - 2 Re\{ x^\dagger A\ket{b} \} + 1,
\label{Ansatz_tree_approach_loss_function}
\end{eqnarray}
while its gradient is defined as
\begin{eqnarray}
\nabla L_R(x)=2A^2x-2A\ket{b}.
\label{Ansatz_tree_approach_loss_function_gradient_overlap}
\end{eqnarray}

The minimization is done by constructing the solution $x$ in the form of an expanding tree,
\begin{eqnarray}
x&=&\alpha_0\ket{b}+\alpha_1U_{v_1}\ket{b}+\alpha_2U_{v_2}U_{v_1}\ket{b}+\dots
\label{x_ansatz_tree_approach}
\\
&\equiv& \alpha_0 |0\rangle + \alpha_1 |1\rangle + \alpha_2 |2\rangle + \dots,
\nonumber
\end{eqnarray}
where $\alpha_k$ are variational parameters. The unitaries $U_{v_1}$, $U_{v_2}$, ... are consecutively determined according to the following iterative procedure:
\begin{enumerate}

\item[0.] At the beginning the set $S$ of the nodes of the tree contains the root $|b\rangle$: $S = \{\ket{b}\}$. At each next step $m$, we perform the following:

\item Find the optimal $x^s=\sum_{j=0}^m\alpha_j\ket{j}$ by optimizing the loss function (\ref{Ansatz_tree_approach_loss_function}) over the parameters $\alpha_0,\dots,\alpha_m$.

\item For each child quantum state $\ket{c} \in C(S) =\{U_1\ket{m},U_2\ket{m},\dots,U_{K_A}\ket{m}\} $ compute the gradient overlap $g=\bra{c}\nabla L_R\left(x^s\right)=2\sum_{j=0}^m\alpha_j\bra{c}A^2\ket{j}-2\bra{c}A\ket{0}$.

\item  Add a new tree node with the largest gradient overlap:  $S\leftarrow S\cup \{\ket{m+1}\}$,  $\ket{m+1}=\mbox{argmax}_{\ket{c}\in C\left(S\right)}|g|$.
\end{enumerate}

\subsection{Efficient ATA for the heat equation}

In Ref. \cite{guseynov2023depth}, it is shown that in the ATA algorithm the computational complexity of a single time step for the heat equation scales as $\mathcal{O}(d^2n^4)$, where $n$ is the number of qubits encoding $\ket{b}$ (logarithm of the matrix size (\ref{A_definition})), and $d$ is ATA tree depth (the number of terms in Eq. (\ref{x_ansatz_tree_approach}) to achieve a given accuracy). It was found that for the sufficiently large grid parameter $c$  the depth of the Anzats tree saturates and does not grow  as the number of qubits $n$ increases. By contrast, the classical algorithms requires $\mathcal{O}(2^n)$ elementary operations for the single time step evolution of the heat equation. 

This exponential speedup of the ATA computational scheme relies on the ability to perform an efficient unitary decomposition of the matrix $A$, see Eq.(\ref{A_definition}). We note that due to periodic boundary conditions the matrix $A$ can be diagonalized by Fourier transform: $M_\mathrm{QFT} A M_\mathrm{QFT}^\dagger = \mbox{diag}\{ \lambda_k^A \}$, 
\begin{equation}
    \lambda_k^A = -c-4\sin^2\left(\frac{\pi k}{2^n}\right),
    \label{A_spectrum}
\end{equation}
where $M_\mathrm{QFT}$ is the Fourier transform unitary matrix.

\begin{figure}[h]
\includegraphics[width=0.38\textwidth]{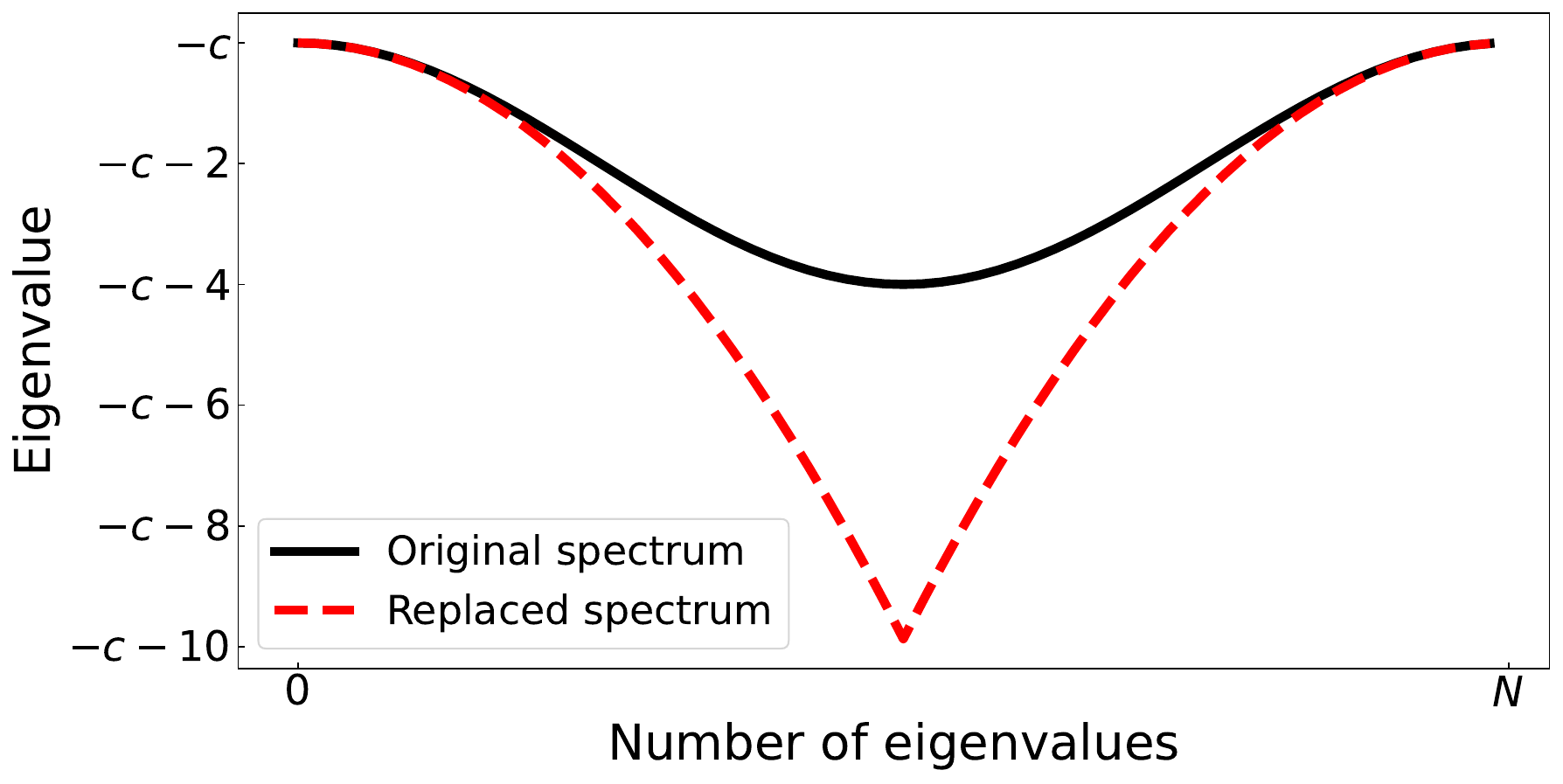}
\caption{The original spectrum of the matrix $A$ and its piecewise quadratic approximation at small wave vectors $2\pi k/2^n$.}
\label{Spectrum_and_substituted_spectrum}
\end{figure}

We key ingredient to the efficient unitary decomposition is to replace the spectrum (\ref{A_spectrum}) by a piecewise quadratic approximation valid at small wave indexes $k$,
\begin{equation}
    \lambda_k^A \to \lambda_k^{\prime A} =  c-\pi^2\left(|\frac{k}{2^{n-1}}-1|-1\right)^2.
    \label{A_fourier_diag_substituted_spectrum}
\end{equation}
This transformation is justified as far as the finite-difference scheme assumes the smallness of the spatial Fourier components in the solution with a large wave vector $2\pi k/2^n$. The switching to the piecewise quadratic spectrum is illustrated in Fig.~\ref{Spectrum_and_substituted_spectrum}.

It was proven in Ref.\cite{guseynov2023depth} that a diagonal matrix with a piecewise quadratic spectrum can be decomposed into a polylogarithmic number of Pauli products. In particular, the matrix $A' = M_\mathrm{QFT}^\dagger \mbox{diag}\{\lambda_k^{\prime A}\} M_\mathrm{QFT}$ can be decomposed into at most $n(n+1)/2$ unitary terms,
\begin{eqnarray}
\begin{gathered}
A^{\prime}=M_\mathrm{QFT}^\dagger\Biggl(
\sum_{i,j}d_{ij}Z_iZ_j+\sum_is_iZ_i+\zeta I
\Biggr)M_\mathrm{QFT}.
\label{A_substituted_decomposition}
\end{gathered}
\end{eqnarray}
The Fourier transform requires $\mathcal{O}(n^2)$ quantum gates and the overall complexity of each unitary term in the decomposition (\ref{A_substituted_decomposition}) is quadratic in the qubit number. Therefore, the decomposition (\ref{A_substituted_decomposition}) satisfies both of the efficiency requirements discussed in Section \ref{sub:linear system}. The error associated with the substitution $A\to A^\prime$ can be reduced by filtering out fast oscillating harmonics in $|b\rangle$.

\subsection{Quantum state generation}

\begin{figure}[!h]
\includegraphics[width=0.89\linewidth]{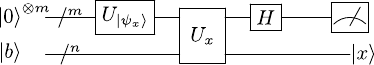}
\caption{The quantum circuit which generates the normalized solution $\ket{x}$. The solution is prepared conditionally on measuring $\ket{0}^{\otimes m}$ in the upper register.}
\label{x_preparation_ansatz_tree_approach_circuit}
\end{figure}

Now we discuss how the quantum state encoding the solution can be efficiently generated for a given tree nodes $S$ and weights $\alpha_i$, see Eq. (\ref{x_ansatz_tree_approach}). This can be done by the quantum circuit shown in  Fig.~\ref{x_preparation_ansatz_tree_approach_circuit}.  The circuit has two qubit registers: (i) $n$-qubit register which contains the solution quantum state, (ii) and an auxiliary $m$-qubit register with $m = \bigl[\log_2(d)\bigr]$. Both registers start from $|0^{\otimes m}\rangle$ and $|0^{\otimes n}\rangle$ states and the $m$-qubit quantum gate $\hat{U}_{|\psi_x\rangle}$ prepares the auxiliary register in the state
\begin{eqnarray}
\ket{\psi_x}=\sum_{i=0}^{m-1}\alpha_i\ket{i},
\label{U_psi_x_ansatz_tree_approach}
\end{eqnarray}
with its amplitudes being the optimized variational parameters of the ATA scheme, see Eq.(\ref{x_ansatz_tree_approach}). Next, the solution register is subjected to the $m$-qubit controlled unitary operation, 
\begin{equation}
    \hat{U}_x = \sum_{j=0}^{m-1} |j\rangle \langle j| \otimes \hat{U}_{v_j} \hat{U}_{v_{j-1}} \dots \hat{U}_{v_1},
    \label{U_x_ansatz_tree_approach}
\end{equation}
where $\hat{U}_{v_j}$ are unitary operators of the Ansatz tree, see Eq.((\ref{x_ansatz_tree_approach})). According to Eq.(\ref{A_substituted_decomposition}) each controlled unitary operation in the decomposition (\ref{U_x_ansatz_tree_approach}) can be constructed through two Fourier transforms (direct and inverse one) acting on the $n$-qubit register with a number of $CZ$-gates in between. Finally, we apply an $m$-qubit Hadamard gate to the auxiliary register and measures its state. The correct solution state is prepared in the $n$-qubit register once the $0^{m}$-bit string is measured. 

Let us analyze the complexity of this preparation procedure.  The unitary $\hat{U}_{|\psi_x\rangle}$ can be decomposed into a circuit with  $\mathcal{O}(2^m)$ depth. The two Fourier transforms require $n(n-1)$ two-qubit quantum gates

At the same time, the overall mean probability of the construction is $1/2^ m$. However, it should be noted that the number of qubits in the upper register is $m={\lceil\log_2d\rceil}$, where $d$ is the tree depth. Thus, the unitary operation $U_{\ket{\psi_x}}$ complexity and the overall probability of the quantum circuit shown in the Fig.~\ref{x_preparation_ansatz_tree_approach_circuit} do not scale exponentially with $n$.

Thus, the circuit depth of the presented scheme of construction of the solution $x$ shown in the Fig.~\ref{x_preparation_ansatz_tree_approach_circuit} scales as $\mathcal{O}(n^2)$ and coincides with the complexity of the Fourier transform. Note that the quantum circuit is a bottleneck for the efficient application of ATA to the heat equation, since we should limit the number of $m\sim poly(\log n)$ to achieve quantum supremacy for two reasons. First, the number of qubits in the upper auxiliary register determines the probability of obtaining the solution, which is an initial state for the next time transition. Second, $U_{\ket{\psi_x}}$  is an exponentially hard operation, so to achieve quantum supremacy $m$ has to be exponentially smaller than $n$.

\section{Successive application of ATA}

\begin{figure}[h]
\includegraphics[width=0.47\textwidth]{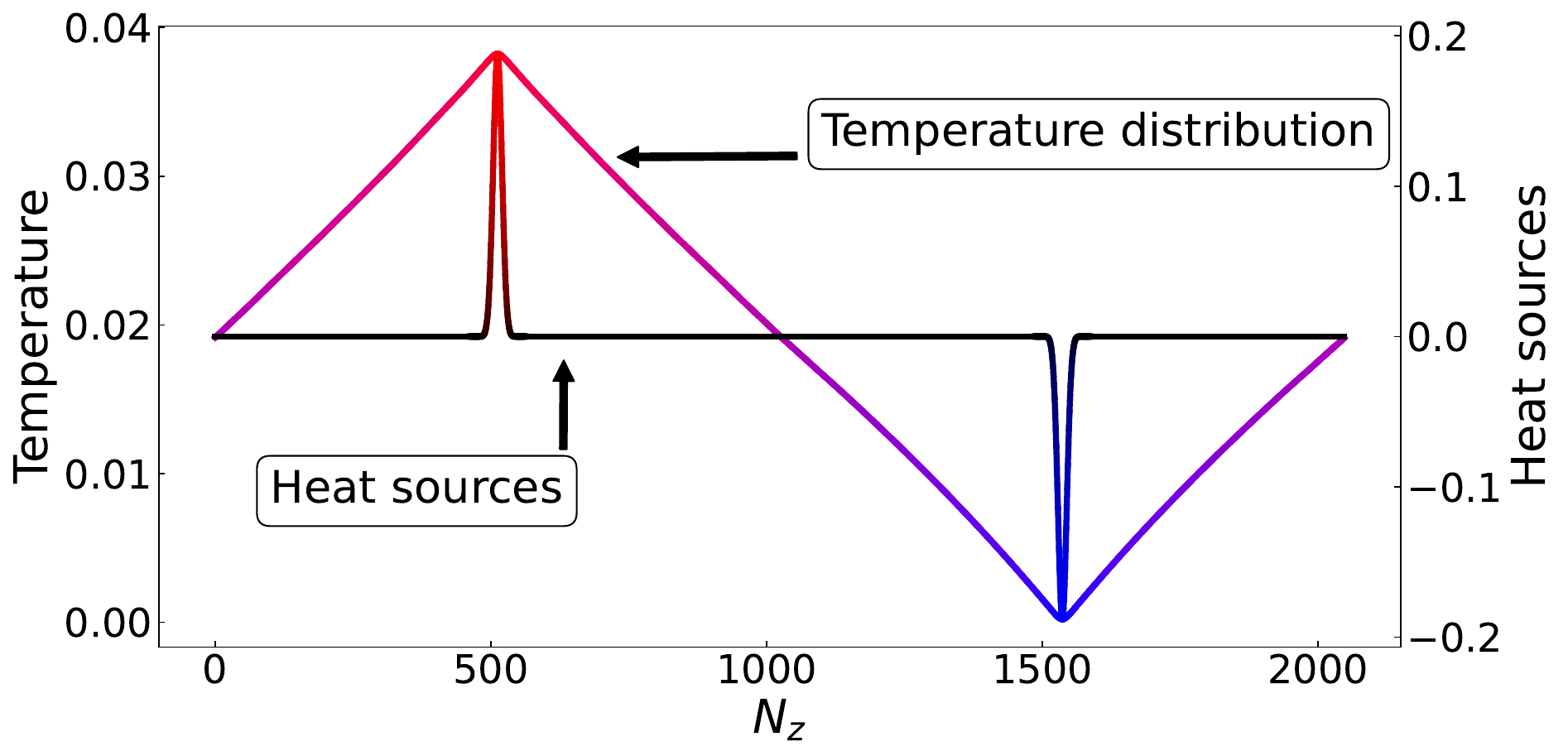}

\caption{Stationary solution for a quasi-singular heater and cooler for $n=11$. The data shown are normalised since ATA yields a solution up to a constant factor.}\label{stationary_famous}
\end{figure}

In the previous section we showed how to implement the starting time transition $t=0 \to \delta t$ using ATA. Our next step is successively apply ATA and find the solution of heat equation at an arbitrary time instant. This requires to add the heat sources $f(z,t)$ function to the current temperature distribution at time $t$ according to the formula (\ref{equation_grid}). The addition of the heat source function is one of the most problematic part of this algorithm. Such a simple action can be very problematic in the quantum computing paradigm since this operation does not correspond to a unitary operation. Let us first address the formula
\begin{eqnarray}
b^\tau_i=x_i^\tau+\gamma f_i^\tau.
\label{b=x+f}
\end{eqnarray}
The coefficient $\gamma$ is a small coefficient that depends on the grid parameter $c$. Up to normalization this equation is analogous to the Eq. (\ref{equation_grid}). We assume that we know efficient unitaries that construct the initial temperature distribution $b^0_i$ as well as the heat sources $f_i^\tau$
\begin{eqnarray}
b^0_i=U_b^0\ket{0}^{\otimes n};\qquad f^\tau_i=U_f^\tau\ket{0}^{\otimes n}.
\label{b_and_funitaries}
\end{eqnarray}
Using ATA solution form (\ref{x_ansatz_tree_approach}) and Eqs. (\ref{b=x+f}), (\ref{b_and_funitaries}) we can construct the sum of unitaries that prepares $b^1$
\begin{eqnarray}
\begin{gathered}
b^1=\left(\gamma U_f^1+\sum_{i=0}^d\alpha_iU^0_{v_i}U^0_{v_{i-1}}\dots U^0_{v_1}U_b^0\right)\ket{0}^{\otimes n}\\ \equiv 
\sum_{i=0}^{d+1}\beta^1_iU_{bi}^1\ket{0}^{\otimes n},
\end{gathered}
\label{b_1_through_unitaries}
\end{eqnarray}
where $d$ is the depth of ATA. Thus, the vector $b^1$ can be constructed on a quantum computer using the quantum circuit shown in Fig.~\ref{x_preparation_ansatz_tree_approach_circuit}. The vector $b^1$ is used as the initial condition for the next time step, so by induction  on the $k$-th time step one gets,
\begin{eqnarray}
b^k&=&\gamma U_f^k \ket{0}^{\otimes n} +
\Bigl[\sum_{i=0}^d\alpha_iU^{k-1}_{v_i}U^{k-1}_{v_{i-1}}\dots U^{k-1}_{v_1} \Bigr] b^{k-1} 
\nonumber\\
&\equiv& \sum_{i=0}^{d_{k}}\beta_i^kU_{bi}^k\ket{0}^{\otimes n}.
\label{b_i_through_unitaries}
\end{eqnarray}
Note that the number of unitarities $U_{bi}^k$ generally grows at most exponentially with $k$: $d_k \leq d\, d_{k-1} +1$. In Section \ref{ch:dropout} we propose the efficient way to avoid this problem by dropping out the low-impact terms in the Eq.~(\ref{b_i_through_unitaries}).

In order to demonstrate the feasibility of our method we apply it to a test model of two quasi-singular stationary heat sources, see Fig.~\ref{stationary_famous}. The established solution with the fidelity over $0.999$ has been observed after $N_\tau=20000$ time steps with the depth tree restricted by $d=35$ and $c=0.1$ at each time transition.  Thus, we conclude that it is possible to achieve an accurate time-developed solution by using the presented unitary decomposition with bounded $d$. Although, we have found \cite{guseynov2023depth} the tree depth saturates with $n$: $d\sim\mathcal{O}(1)$ the quantum supremacy is preserved even with $d\sim \mathcal{O}(poly (n))$.

\section{Accuracy analysis for the time dependent problem}\label{s:accuracy for time dependent}

In this section we investigate how the accuracy of the solution changes with time and analyse how the depth of the Ansatz tree develops. We numerically study the required depth of the Ansatz tree needed to achieve a given accuracy. In particular, we investigate the tree depth dependence on the smoothness of the initial temperature distribution $\chi(z)$ and the heat source $f(z,t)$ functions, as well as on the grid parameter $c$. 

The number of qubits $n$ determines the spatial grid size $\delta z$ which should be sufficiently small to catch all scales of the temperature distribution changes. On the other hand, at a fixed $n$ the grid parameter $c$ is determined by the time step size: $c\sim1/\delta t$. Therefore, at small $c$ the two successive in time temperature distributions become very dissimilar to each other that results in a large tree depth to get the next time step solution with a high fidelity.

Below, we define a smoothness of the heat sources and the initial temperature distribution as the maximal degree $\mathcal{G}$ of the Chebyshev polynomials $T_i$ entering into its expansions, 
\begin{eqnarray}
\begin{gathered}
f(z,t)=\sum_{i=0}^{\mathcal{G}^f_t}\sum_{j=0}^{\mathcal{G}^f_Z}h_{ij}T_i(2t-1)T_j(2z-1),\\
\chi(z)=\sum_{i=0}^{\mathcal{G}^\chi_Z}g_iT_i(2z-1),
\label{smoothness}
\end{gathered}
\end{eqnarray}
where $z,t \in[0,1]$.

\begin{figure}[h]
\includegraphics[width=0.45\textwidth]{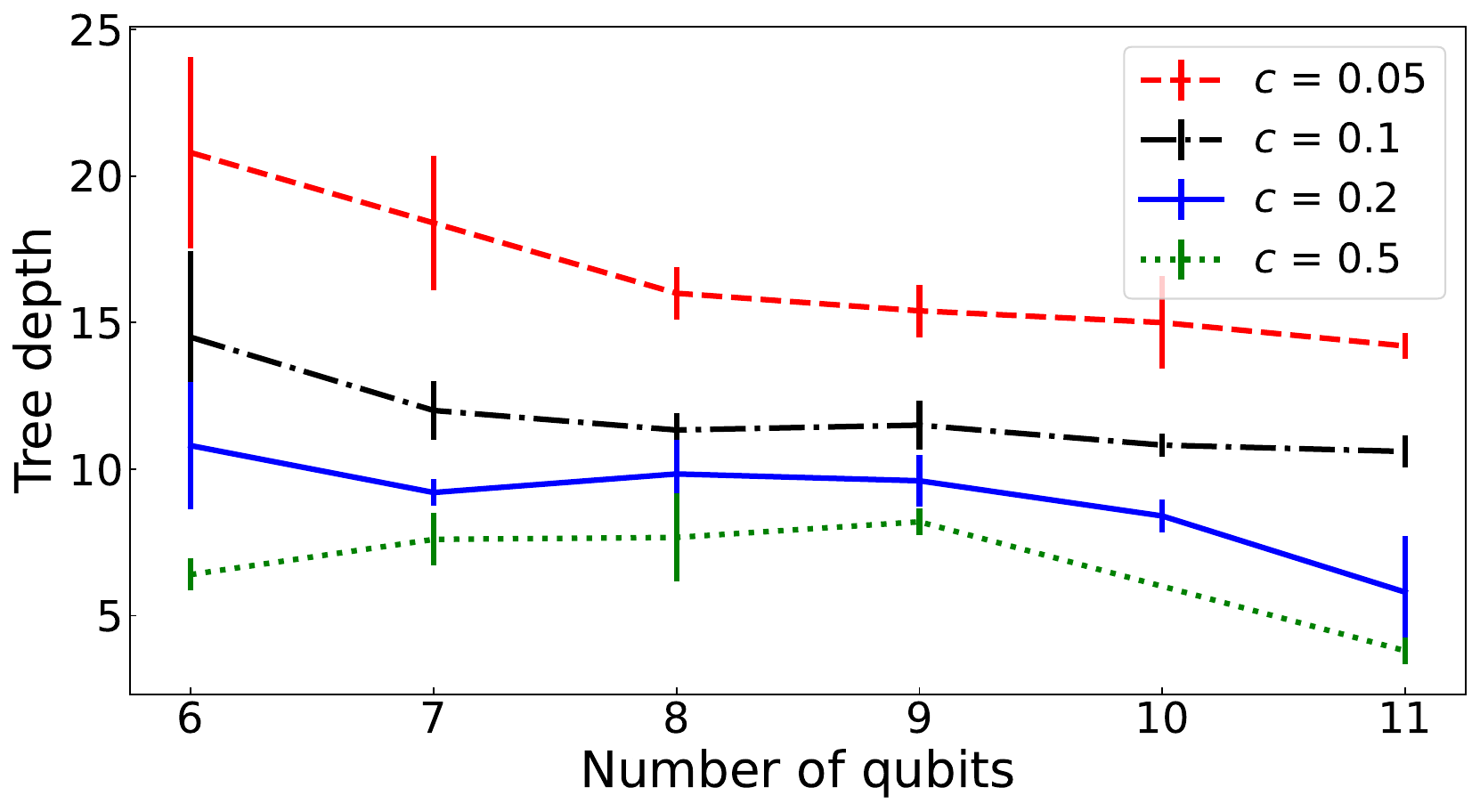}

\caption{The minimal depth of the Ansatz tree needed to achieve the fidelity $0.99$ after $200$ time steps as a function  of the qubits number at different values of the grid parameters $c$. Each point on the graph corresponds to the average over 10 random heat sources and initial temperature distributions with $\mathcal{G} = 20$. Error bars represent the standard deviation.}\label{smoothness_c}
\end{figure}

Next, we numerically determine the minimal depth of the Ansatz tree 
for each time transition required to get a given fidelity of the solution after $N_\tau \gg 1$ time steps.  In Fig. \ref{smoothness_c} we plot the average minimal depth of the tree required to get the fidelity $0.99$ after $N_\tau = 200$ sequential time steps at different values of the grid parameter $c$ and the fixed smoothness $\mathcal{G} =\mathcal{G}_Z^\chi =\mathcal{G}_Z^f =\mathcal{G}_t^f = 20$ and  random $\chi(z)$ and $f(z,t)$ with uniformly distributed coefficients $g_i$ and $h_{ij}$. We also assume that 
\begin{eqnarray}
\int_0^1\int_0^1 |f(z,t)|dzdt\sim\int_0^1|\chi(z)|dz,
\label{smoothness_norm}
\end{eqnarray}
implying that the amount of heat generated by the source $f(z,t)$ during evolution is of the order of the heat initially contained in the system. The latter condition implies that both the initial heat and the heat sources have the same order impact on the final temperature distribution $b^{t=1}$. One can see, that a fixed qubits number $n$ the smaller $c$ (i.e. larger time step size $\delta t$) results in a larger tree size, as expected, and at large $c$ the tree depth gets saturated. Similarly, increasing the qubits number $n$ at a fixed $c$ one effectively reduces the time step size $\delta t$ and thus the required tree depth reduces as well.

\begin{figure}[h]
\includegraphics[width=0.45\textwidth]{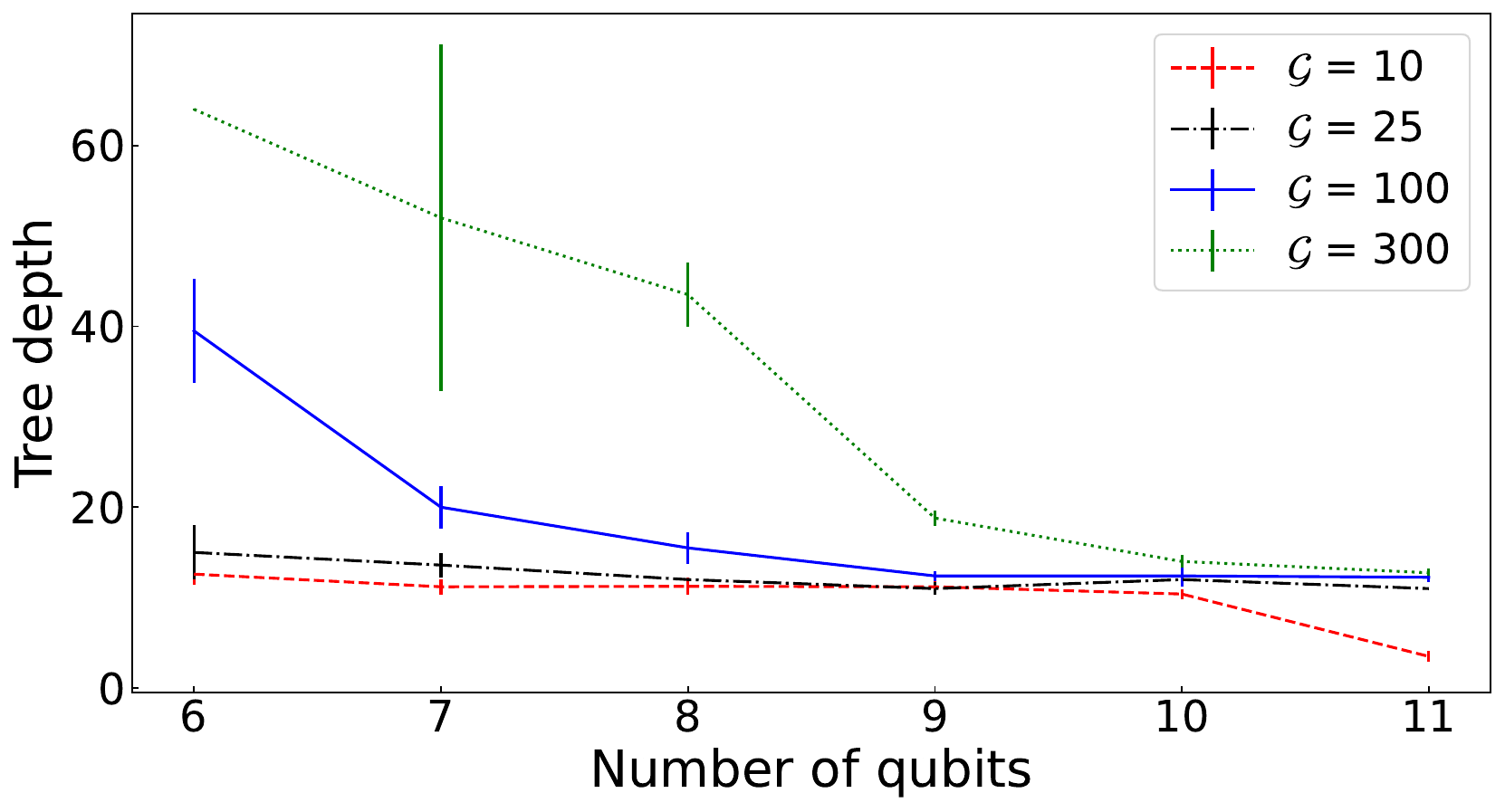}

\caption{
The minimum depth of the Ansatz tree needed to achieve the fidelity $0.99$ after $200$ time steps as a function of the qubits number $n$ at different values of the smoothness parameter $\mathcal{G}$ and the fixed grid parameter $c=0.1$. Each point on the graph is an average over 10 random heat sources and initial temperature distributions. Error bars represent the standard deviation.}\label{smoothness_poly}
\end{figure}

The similar behaviour of the tree depth is observed with the change of the smoothness parameter $\mathcal{G}$, see Fig.~\ref{smoothness_poly}. At a constant $c=0.1$ and the fixed qubits number $n$ the more irregular initial heat distribution and heat sources requires a larger tree depth. For example, at $\mathcal{G} = 300$ and $n=6$ the tree depth approaches its maximum value $d=2^6$ as far as this number of qubits is insufficient to resolve the fine structure of the $\chi(t)$ and $f(z,t)$ functions, and at each next time steps its values looks almost random and uncorrelated with the previous time instant.  Therefore, at any random choice of the $\chi(t)$ and $f(z,t)$ the tree depth is equal $2^6$ and no error bar is present in this case, see Fig.~\ref{smoothness_poly}. Increasing the number of qubits one resolves all spatial and temporal scales at the fixed $\mathcal{G}$ and the tree depth decreases. In general, at a fixed time step $\delta t$ and a large qubits number $2^n \gg \mathcal{G}$ the tree depth should scale as $d \sim \mathcal{O}(\mathcal{G})$. Indeed, the complexity (i.e. number of quantum gates) to prepare a $n$-qubit state with $\mathcal{G} \ll 2^n$ different amplitudes in computational basis scales as  $\mathcal{O}(\mathcal{G})$. However, one can clearly see, that at large $n$ the tree depth is saturated to a constant value almost independent on $\mathcal{G}$. We argue, however, that at the fixed $c$ the time step size exponentially decreases with the qubits number: $\delta t \sim 2^{-2n}$ and therefore the change $f(z,t)\delta t$ becomes negligible and two nearest temperature distributions $b^{t}$ and $b^{t+\delta t}$ are almost the same. As a result, the corresponding Ansatz tree for the transition $b^{t} \to b^{t+\delta t}$ requires $d \ll \mathcal{O}(\mathcal{G})$. However, the more accurate investigation  of the depth scaling law requires numerical simulation with a larger number of qubits.

\section{Repetitive tree nodes}\label{s:repetitive}

In this section we study how the Ansatz tree nodes develop with time.  We find a stationary cluster of tree nodes which contains unitary operations that are preserved during time evolution at all time instants. This cluster can be exploited for reducing the computational costs during the learning of the Ansatz tree  at each time step $\tau \to \tau+1$ as one can start the tree growth already from the cluster nodes,
\begin{eqnarray}
x^{\tau+1}_{start}=\Bigl[\sum_i\gamma^{start}_iU^{e-p}_{i}\Bigr] b^\tau,
\label{smoothness}
\end{eqnarray}
rather than from the state $b^\tau$, see Eq.(\ref{b_i_through_unitaries}). The nodes of the stationary cluster can be found during first several time steps of the ATA, and the weighting coefficients $\gamma_i^{start}$ are determined through minimization of the loss function, see Section \ref{s:prelim}. 

\begin{figure}[h]
\includegraphics[width=0.8\linewidth]{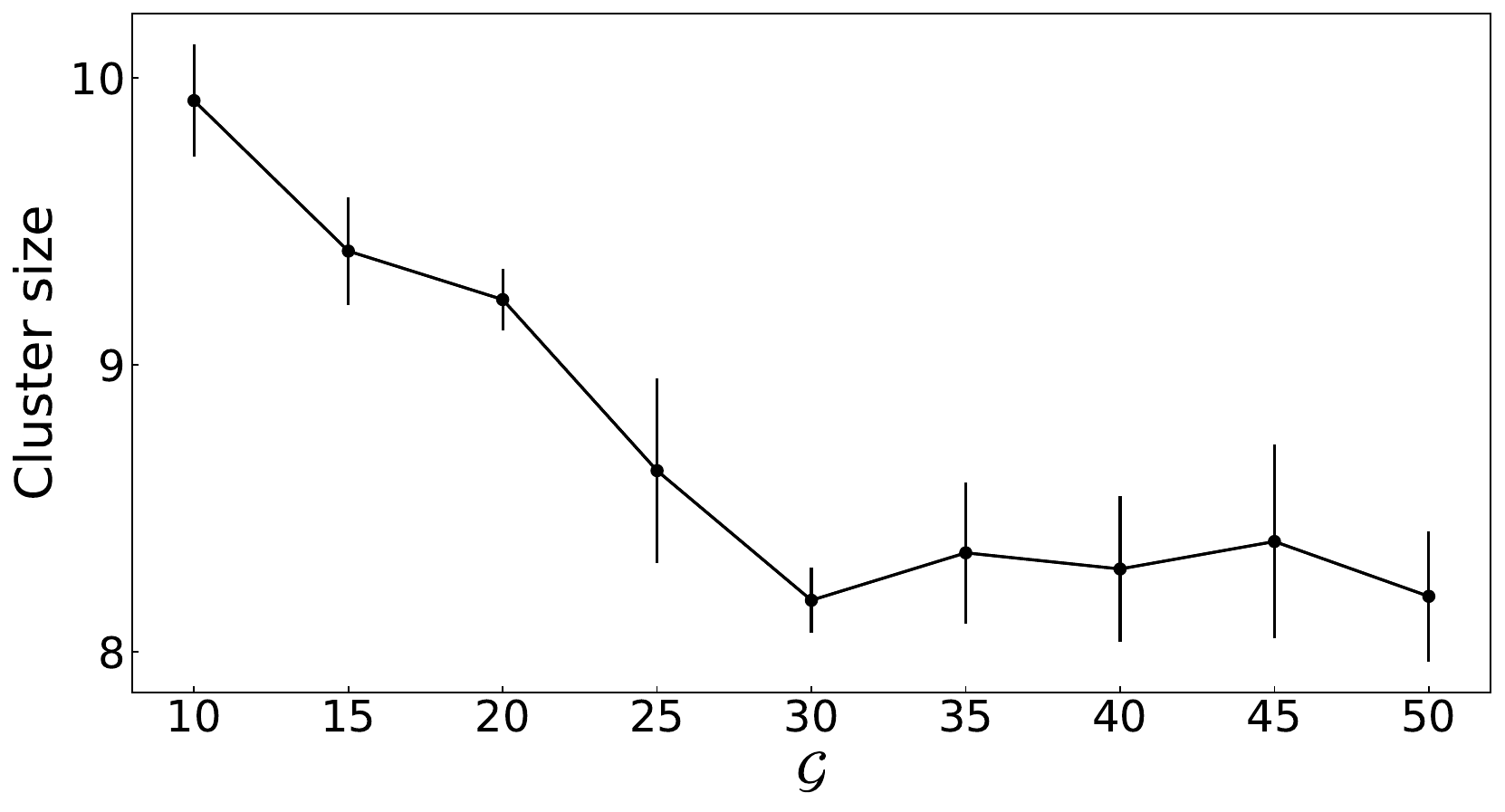}

\caption{The dependence of the stationary cluster size on the smoothness parameters $\mathcal{G}$ for $n=11$ qubits and the grid parameter $c=0.1$.  Each point on the plot corresponds to the average of $50$ random heat sources and initial temperature distributions.}
\label{cluster_poly}
\end{figure}

First, we study the dependence of the stationary cluster size on the smoothness parameter $\mathcal{G}$, see Eq. (\ref{smoothness}). At each numerical experiment we randomly choose $\chi(z)$ and $f(z,t)$ function and execute a large number $M=1000$ of time steps each restricted to the depth $d=30$. Next, we determine an overlap between the nodes of $M$ Ansatz trees which forms the stationary cluster. Fig.~\ref{cluster_poly} shows how the average stationary cluster size changes as the smoothness $\mathcal{G}$ increases. The stationary cluster size saturates with $\mathcal{G}$ as far as the temperature distributions for successive time steps becomes dissimilar for a large $\mathcal{G}$. 

\begin{figure}[!h]
\includegraphics[width=0.8\linewidth]{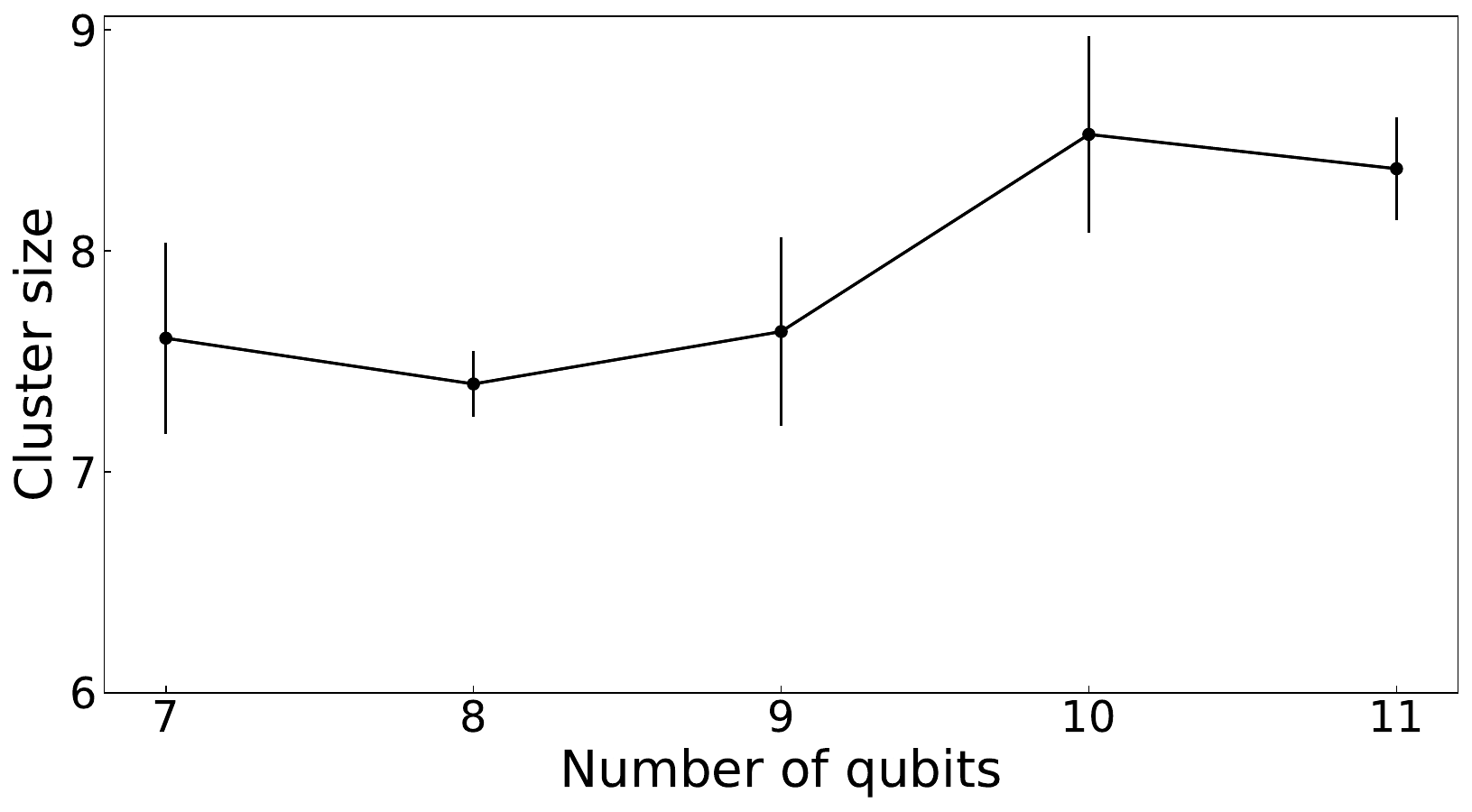}

\caption{The dependence of the stationary cluster size on the qubit number for the random and independent initial temperature distributions. Each point on the plot corresponds to an average over $50$ samples randomly selected according to the Haar measure. As a stop criterion we use the depth $d=50$.}
\label{cluster_random_b}
\end{figure}

We argue, that the nature of the stationary cluster has nothing to do with the specific form of the heat source function $f(z,t)$. In order to check this hypothesis we do another simulation where we randomly choose the initial temperature distribution and construct the Ansatz tree for a single time transition. Comparing the trees from the different samples we do find the same stationary cluster as in the previous simulation with the large $\mathcal{G}$.  The cluster size does not growth with the qubits number, see Fig.~\ref{cluster_random_b}. It follows that the nature of the stationary cluster is determined only by the heat equation itself rather than by a certain form of $\chi(z)$ and $f(z,t)$ functions.

\section{Dropout of the expanding Ansatz tree}
\label{ch:dropout}

In Section \ref{s:accuracy for time dependent} we faced with the problem of the exponential expansion of the Ansatz tree with time. This behaviour destroys the quantum supremacy as long as: (i) the quantum circuit Fig.~\ref{x_preparation_ansatz_tree_approach_circuit} which prepares the solution state $|b^\tau\rangle$ has an exponential depth, (ii) due to the probabilistic character of the preparation procedure the probability to prepare the solution state becomes exponentially small and the learning procedure of the Ansatz tree becomes an exponentially hard problem.

Here we suggest the strategy which heals that problem by limiting the number of unitaries that produce $\ket{b^\tau}$, see Eq. (\ref{b_i_through_unitaries}). After each time step we leave only $D_\mathrm{cut}$ unitaries with the largest absolute weights $\kappa^\tau_i$,
\begin{eqnarray}
    b^\tau=M_{\mbox{QFT}}^\dagger
    \left(\sum_{i=0}^{D_\mathrm{cut}}\kappa^\tau_i\Pi_{\kappa s^\tau_i}\right)
    M_{\mbox{QFT}}\ket{0}^{\otimes n},
\label{b_tau_drop}
\end{eqnarray}
and drop out the remaining nodes. As a result, at any time step the depth of the Ansatz tree is limited by $D_\mathrm{cut}$. In order to numerically model the successive ATA procedure with $D_\mathrm{cut}$ tree size we assume that the heat sources and initial temperature distribution are given in the same form as the solution itself, 
\begin{eqnarray}
\begin{gathered}
    f^\tau=M_{\mbox{QFT}}^\dagger
    \left(\sum_{i=0}^{d^\tau_f}h^\tau_i\Pi_{hs^\tau_i}\right)
    M_{\mbox{QFT}}\ket{0}^{\otimes n};\\
    b^0=M_{\mbox{QFT}}^\dagger
    \left(\sum_{i=0}^{d^0_b}\kappa^0_i\Pi_{\kappa s^0_i}\right)
    M_{\mbox{QFT}}\ket{0}^{\otimes n},
\end{gathered}
\label{b_f_drop}
\end{eqnarray}
where $\Pi_{hs^\tau_i}$ and $\Pi_{\kappa s^0_i}$ are Pauli products of $I$ and $Z$ operators only. This representation is general, as proven in Appendix \ref{app: arbitrary with I and Z},  however, the similar simulation of the dropout procedure can be performed for any representation of $f$ and $b^0$.

\begin{figure}[h]
\includegraphics[width=\linewidth]{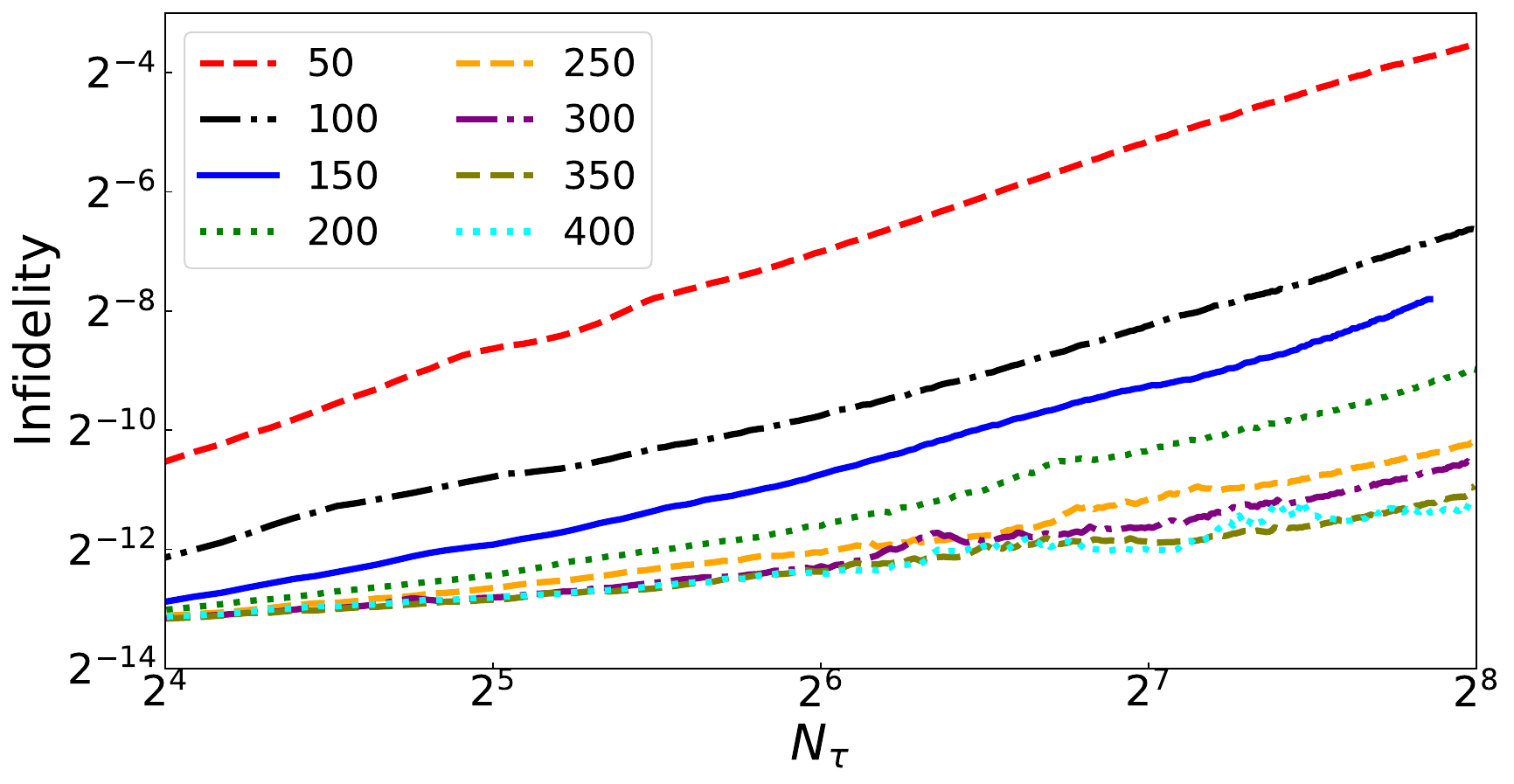}
\caption{
The infidelity of accurate and sequential dropped out solutions depending on the number of time steps (the number of dropout applications) for different $D_{cut}$ and $n=11$. 
}
\label{drop_out}
\end{figure}

Fig.~\ref{drop_out} shows the efficiency of the dropout algorithm described above. In our numerical simulations for $n=11$ we use random initial temperature distributions and heat sources according to Eq.~(\ref{b_f_drop}), the coefficients $h_i^\tau$ change slowly in time according to a random polynomial of degree $\mathcal{G}=20$. Initially, the coefficients $h_i^\tau$ and $\kappa_i^0$ are randomly chosen from the uniform distribution, the number of Pauli products in the temperature distributions does not exceed $D_{cut}$. For each time step we use the grid parameter $c=2$ and the tree depth $d=50$, this choice of parameters is justified by the non-smoothness of the temperature distribution, which is consistent with the Ref. \cite{guseynov2023depth}. Each curve in the figure is generated by averaging over 10 different initial temperature distributions and heat sources that satisfy the condition (\ref{smoothness_norm}).

Fig.~\ref{drop_out} shows that the error generated by the dropping of of the overall algorithm scales linearly $\epsilon \sim \mathcal{O}(N_\tau)$ with presented strategy of dropping Pauli products. We compare this estimation with classical roundoff error \cite{press2007numerical} which is a lower bound for computation with float numbers. For tridiagonal matrix algorithm this type of error scales as $\epsilon_{round} \sim \sqrt{N_\tau}$. On the other hand, the condition number \cite{belsley2005regression} error arises each time we solve system (\ref{Ax=b}; this error provides an upper bound for the classical error estimation. The work \cite{guseynov2023depth} estimates the contribution of this type of error by $\mathcal{O}((c+4)^{N_\tau})$. Thus, our result demonstrates the quadratically higher error scaling in comparison with the roundoff error and exponentially smaller one in comparison with the upper bound estimation.



\section{Comparison with the HHL algorithm}\label{s:HHL}

In this section, we compare the performance of the ATA and the HHL algorithm \cite{hhl} in solving the heat equation. We start our consideration with a fault-tolerance case. The HHL algorithm is probabilistic; it returns the solution when its ancilla qubit is measured in the state $|1\rangle$ with some probability $p<1$. The solution of the heat equation (\ref{Thermal_conductivity_equation}) requires $N_\tau$ times applications of the HHL algorithm to the linear system (\ref{equation_grid}) at each time step, where the initial condition at the $i$th time step involves the solution of the preceding $i\!-\!1$th time step, see Eq.(\ref{equation_grid}). This means that the correct solution of the heat equation at the final time step is obtained only with an exponentially small probability $\sim (1-p)^{N_\tau} \approx \exp(-pN_\tau)$ implying that the ancilla qubit is measured in the state $|1\rangle$ at the all sequential applications of the HHL algorithm. Thus the probabilistic nature of the HHL algorithm significantly restricts its application for the heat equation problem even in the fault-tolerant case.

In the ansatz tree approach one also generates the solution of the heat equation in the probabilistic manner as shown in the Fig. \ref{x_preparation_ansatz_tree_approach_circuit}, 
where the solution is returned only when $m$ ancilla qubits are all measured in the state $|0\rangle$ with the probability $\sim 1/d$ where $d$ is the anzatz depth. The important difference with the HHL approach is that we build up the ansatz tree in a classical manner: once we learned the ansatz tree for the time step $t$ we need not repeat creation of all the previous temperature distributions while constructing the initial condition for the next time step. This allows us to circumvent the exponentially decay of the probability to get the correct answer: the probabilistic overhead in the number of runs of the circuit Fig. \ref{x_preparation_ansatz_tree_approach_circuit} results in  the penalty factor $\sim N_\tau d$ which scales linearly with $N_\tau$ in contrast to the exponential penatly factor $\exp(N_\tau p)$ in the number of runs of the HHL algorithm.  

The construction of the initial condition for the next time step can be done efficiently, provided the heat source function $f^\tau$ can be generated at each time by the known ansatz tree applied to the initial state $\ket{0}^{\otimes n}$, see Fig.~\ref{x_preparation_ansatz_tree_approach_circuit}. Indeed, as follows from the Eqs. (\ref{equation_grid}), (\ref{x_ansatz_tree_approach}), (\ref{b_f_drop}) the initial condition for the time step $\tau+1$ is given by,
\begin{eqnarray}
\begin{gathered}
b^\tau=\frac{(\delta z)^2}{a^2}f^\tau+c\,x^\tau=\frac{(\delta z)^2}{a^2}M_{\mbox{QFT}}^\dagger\left(\sum^{K^t_f}_{i=0}h_i^\tau\Pi_{hs^\tau_i}\right)\\\times 
M_{\mbox{QFT}}\ket{0}^{\otimes n}+c\,M_{\mbox{QFT}}^\dagger\left(\sum^{K^t_x}_{i=0}\kappa_i^{\tau-1}\Pi_{\kappa s^{\tau-1}_i}\right) \\\times
M_{\mbox{QFT}}\, b^{\tau-1}=\dots\\
=M_{\mbox{QFT}}^\dagger\left(\sum^{K_b^0}_{i=0}\kappa^0_i\Pi_{\kappa s^0_i}\right)M_{\mbox{QFT}}\,\ket{0}^{\otimes n}.
\label{initial_cond_ith_forming}
\end{gathered}
\end{eqnarray} 

Next, let us discuss the noise sensitivity of the variational ansatz tree approach and the HHL approach due to the gate errors. The most important parameter in this respect is the number of two qubits gates required to build the variational ansatz, see Fig. \ref{x_preparation_ansatz_tree_approach_circuit}, and the HHL quantum circuit. In the former case the majority of the two-qubit gates come from the Fourier transform which results in $O(n^2)$ two-qubit gates complexity of the variational ansatz. The HHL quantum circuit comprises three different blocks: i) a phase estimation block (PEA), ii) a reciprocal eigenvalue block (REV) and iii) an inverse phase estimation block (iPEA). The PEA block decomposes the initial vector $\vec{b}^0$ into eigensystem of the square matrix $A$, see Eq.(\ref{A_definition}): $|0\rangle |\vec{b}^0\rangle \to \sum_\lambda \beta_\lambda^0 |\lambda\rangle |\vec{a}_\lambda\rangle$ with $A \vec{a}_\lambda = \lambda \vec{a}_\lambda$. For the specific heat equation case the PEA and iPEA blocks can be implemented with $O(n^3)$ number of two-qubit gates. The IEV block makes the reciprocal function transformation of the eigenvalue register and the ancilla qubit: $|0\rangle |\lambda\rangle \to (\sqrt{1-c^2/\lambda^2} |0\rangle + c/\lambda |1\rangle)|\lambda\rangle$. In general, for an arbitrary spectrum $\{\lambda\}$ this transformation requires $O(\exp(n))$ quantum gates, although for the quadratic spectrum, see Eq.(\ref{A_fourier_diag_substituted_spectrum}), one possibly can compose an approximate REV operator with polylog complexity $O(n^p)$ that requires a separate study. Therefore, one can conclude that HHL approach requires at least $O(n^3)$ two-qubit gates that makes it more sensitive to the gate noise.

Now we turn to the complexity of the construction of the initial state $\vec{b
}^0$ which was not considered in the previous sections. The construction of an arbitrary state vector is an exponentially hard problem in general. In order to have quantum advantage of the overall algorithm, an efficient way for generation of the initial state $b^0$ should be introduced. For the ansatz tree approach considered here any generation procedure which has $O(poly(n))$ complexity guarantees an exponential speedup over the classical algorithm \cite{guseynov2023depth}. We argue, that such a requirement corresponds to a sufficiently smooth initial temperature distribution function \cite{grover2002creating}. We interpret this physically meaningful restriction as a limit on the number of non-zero low-frequency harmonics in the Fourier decomposition of the initial temperature distribution.

\section{Conclusions}\label{s:conclusion}

The paper studies the realization of the quantum variational Ansatz tree approach for numerical solution of the heat equation. Using a finite difference scheme, the problem is reduced to the sequential solution of a system of linear equations. After each time transition, the solution of Eq.~(\ref{Ax=b}) must be summed with the distribution of heat sources. In this work, we have proposed ideas how to efficiently realise these two important steps in order to achieve quantum acceleration for large systems. 

The addition of two quantum states on a quantum computer is, generally speaking, a difficult task, but with the help of a special quantum circuit to construct the solution shown in Fig.~\ref{x_preparation_ansatz_tree_approach_circuit} we find a way to implement this  complicated step. This construction approach relies on Ansatzes to create quantum states that need to be added. The obtained stationary solution for two quasi-point thermal sources depicted in Fig.~\ref{stationary_famous} shows that this scheme of addition of quantum states is effective and allows us to reproduce an exact numerical solution.

We investigated Ansatz trees for each time transition and found that it is always possible to identify a repetitive nodes cluster whose size depends on the smoothness of the temperature distribution. Thus, we propose to start the construction of the solution vector (\ref{x_ansatz_tree_approach}) with a cluster of repetitive nodes, which allows us to reduce the computational complexity of the algorithm. We also determined the dependencies of the optimal choice of the stopping parameter of ATA to achieve a given accuracy. These dependencies behave in an intuitive way, which is consistent with the definition of the accuracy of the finite-difference scheme \cite{higham2002accuracy} and the analysis of the conditioning number of the matrix \cite{belsley2005regression}. We note that the Ansatz tree depth does not depend on the matrix size at fixed smoothness and finite-difference mesh partitioning parameters, which is a necessary condition of achieving a quantum acceleration for large systems.

The quantum circuit shown in Fig~\ref{x_preparation_ansatz_tree_approach_circuit} for constructing the ATA solution is probabilistic and also assumes the construction of an arbitrary quantum state. These two facts impose a significant limitation on the auxiliary register of this quantum circuit when operating in the quantum acceleration regime. Thus, the condition for achieving quantum acceleration is the polylogarithmic dependence of the number of auxiliary qubits on the logarithm of the matrix size of the system to be solved (the number of qubits in the main register). In this paper, we have shown that when the size of the auxiliary register is limited, the solution suffers only a small, approximately linear loss in accuracy, which is entirely determined by the size of the auxiliary upper register in Fig.~\ref{x_preparation_ansatz_tree_approach_circuit}. This result can be explained by the nature of the diffusion equation, which forgets the initial conditions after some time. This property allowed us to demonstrate in our simulations for $n=11$ qubits the possibility of providing quantum acceleration in a fault-tolerant regime.

We compared the complexity of ATA with other promising quantum algorithms, in particular the HHL algorithm for solving systems of linear equations. We conclude that the iterative method for solving linear differential equations is not efficient for this non-variational probabilistic algorithm because the solution at the $\tau$-th time step is the initial condition for the $\tau +1$-th step, which generates an exponential decay of the probability of success. On the other hand, ATA does not have this drawback, while restricting the auxiliary register for solution construction, the solution at each step has approximately the same probability of construction. Moreover, we have evaluated the circuit depths for the implementation of the HHL algorithm, one can conclude that the HHL approach requires at least $O(n^3)$ two-qubit gates; however, ATA requires $O(n^2)$ such gates, which makes it more robust to gate noise.

\section{Acknowledgments}\label{s:acknowledgments}
W. V. P. acknowledges support from the RSF grant No. 23-72-30004 (https://rscf.ru/project/23-72-30004/).

\bibliography{references}

\begin{thebibliography}{46}
\expandafter\ifx\csname natexlab\endcsname\relax\def\natexlab#1{#1}\fi
\expandafter\ifx\csname bibnamefont\endcsname\relax
  \def\bibnamefont#1{#1}\fi
\expandafter\ifx\csname bibfnamefont\endcsname\relax
  \def\bibfnamefont#1{#1}\fi
\expandafter\ifx\csname citenamefont\endcsname\relax
  \def\citenamefont#1{#1}\fi
\expandafter\ifx\csname url\endcsname\relax
  \def\url#1{\texttt{#1}}\fi
\expandafter\ifx\csname urlprefix\endcsname\relax\def\urlprefix{URL }\fi
\providecommand{\bibinfo}[2]{#2}
\providecommand{\eprint}[2][]{\url{#2}}

\bibitem[{\citenamefont{Pour-El and Richards}(1982)}]{FEYNMAN}
\bibinfo{author}{\bibfnamefont{M.}~\bibnamefont{Pour-El}} \bibnamefont{and}
  \bibinfo{author}{\bibfnamefont{I.}~\bibnamefont{Richards}},
  \bibinfo{journal}{International Journal of Theoretical Physics}
  \textbf{\bibinfo{volume}{21}}, \bibinfo{pages}{553} (\bibinfo{year}{1982}).

\bibitem[{\citenamefont{Shor}(1994)}]{math_apply}
\bibinfo{author}{\bibfnamefont{P.~W.} \bibnamefont{Shor}}, in
  \emph{\bibinfo{booktitle}{Proceedings 35th annual symposium on foundations of
  computer science}} (\bibinfo{organization}{Ieee}, \bibinfo{year}{1994}), pp.
  \bibinfo{pages}{124--134}.

\bibitem[{\citenamefont{Shor}(1999)}]{Shor}
\bibinfo{author}{\bibfnamefont{P.~W.} \bibnamefont{Shor}},
  \bibinfo{journal}{SIAM Review} \textbf{\bibinfo{volume}{41}},
  \bibinfo{pages}{303} (\bibinfo{year}{1999}).

\bibitem[{\citenamefont{Grover}(1996)}]{grover}
\bibinfo{author}{\bibfnamefont{L.~K.} \bibnamefont{Grover}}, in
  \emph{\bibinfo{booktitle}{Proceedings of the twenty-eighth annual ACM
  symposium on Theory of computing}} (\bibinfo{year}{1996}), pp.
  \bibinfo{pages}{212--219}.

\bibitem[{\citenamefont{Ambainis}(2004)}]{q_search}
\bibinfo{author}{\bibfnamefont{A.}~\bibnamefont{Ambainis}},
  \bibinfo{journal}{ACM SIGACT News} \textbf{\bibinfo{volume}{35}},
  \bibinfo{pages}{22} (\bibinfo{year}{2004}).

\bibitem[{\citenamefont{Jin and Liu}(2023)}]{jin2023quantum}
\bibinfo{author}{\bibfnamefont{S.}~\bibnamefont{Jin}} \bibnamefont{and}
  \bibinfo{author}{\bibfnamefont{N.}~\bibnamefont{Liu}},
  \bibinfo{journal}{arXiv preprint arXiv:2304.02865}  (\bibinfo{year}{2023}).

\bibitem[{\citenamefont{Herman et~al.}(2023)\citenamefont{Herman, Googin, Liu,
  Sun, Galda, Safro, Pistoia, and Alexeev}}]{herman2023quantum}
\bibinfo{author}{\bibfnamefont{D.}~\bibnamefont{Herman}},
  \bibinfo{author}{\bibfnamefont{C.}~\bibnamefont{Googin}},
  \bibinfo{author}{\bibfnamefont{X.}~\bibnamefont{Liu}},
  \bibinfo{author}{\bibfnamefont{Y.}~\bibnamefont{Sun}},
  \bibinfo{author}{\bibfnamefont{A.}~\bibnamefont{Galda}},
  \bibinfo{author}{\bibfnamefont{I.}~\bibnamefont{Safro}},
  \bibinfo{author}{\bibfnamefont{M.}~\bibnamefont{Pistoia}}, \bibnamefont{and}
  \bibinfo{author}{\bibfnamefont{Y.}~\bibnamefont{Alexeev}},
  \bibinfo{journal}{Nat Rev Phys} pp. \bibinfo{pages}{1--16}
  (\bibinfo{year}{2023}).

\bibitem[{\citenamefont{Baiardi et~al.}(2023)\citenamefont{Baiardi, Christandl,
  and Reiher}}]{baiardi2023quantum}
\bibinfo{author}{\bibfnamefont{A.}~\bibnamefont{Baiardi}},
  \bibinfo{author}{\bibfnamefont{M.}~\bibnamefont{Christandl}},
  \bibnamefont{and} \bibinfo{author}{\bibfnamefont{M.}~\bibnamefont{Reiher}},
  \bibinfo{journal}{ChemBioChem} \textbf{\bibinfo{volume}{24}},
  \bibinfo{pages}{e202300120} (\bibinfo{year}{2023}).

\bibitem[{\citenamefont{Hassija et~al.}(2020)\citenamefont{Hassija, Chamola,
  Goyal, Kanhere, and Guizani}}]{hassija2020forthcoming}
\bibinfo{author}{\bibfnamefont{V.}~\bibnamefont{Hassija}},
  \bibinfo{author}{\bibfnamefont{V.}~\bibnamefont{Chamola}},
  \bibinfo{author}{\bibfnamefont{A.}~\bibnamefont{Goyal}},
  \bibinfo{author}{\bibfnamefont{S.~S.} \bibnamefont{Kanhere}},
  \bibnamefont{and} \bibinfo{author}{\bibfnamefont{N.}~\bibnamefont{Guizani}},
  \bibinfo{journal}{{IET} Quant. Commun.} \textbf{\bibinfo{volume}{1}},
  \bibinfo{pages}{35} (\bibinfo{year}{2020}).

\bibitem[{\citenamefont{Guseynov et~al.}(2023)\citenamefont{Guseynov, Zhukov,
  Pogosov, and Lebedev}}]{guseynov2023depth}
\bibinfo{author}{\bibfnamefont{N.}~\bibnamefont{Guseynov}},
  \bibinfo{author}{\bibfnamefont{A.}~\bibnamefont{Zhukov}},
  \bibinfo{author}{\bibfnamefont{W.}~\bibnamefont{Pogosov}}, \bibnamefont{and}
  \bibinfo{author}{\bibfnamefont{A.}~\bibnamefont{Lebedev}},
  \bibinfo{journal}{Phys. Rev. A} \textbf{\bibinfo{volume}{107}},
  \bibinfo{pages}{052422} (\bibinfo{year}{2023}).

\bibitem[{\citenamefont{Linden et~al.}(2022)\citenamefont{Linden, Montanaro,
  and Shao}}]{linden2022quantum}
\bibinfo{author}{\bibfnamefont{N.}~\bibnamefont{Linden}},
  \bibinfo{author}{\bibfnamefont{A.}~\bibnamefont{Montanaro}},
  \bibnamefont{and} \bibinfo{author}{\bibfnamefont{C.}~\bibnamefont{Shao}},
  \bibinfo{journal}{Commun. Math. Phys.} \textbf{\bibinfo{volume}{395}},
  \bibinfo{pages}{601} (\bibinfo{year}{2022}).

\bibitem[{\citenamefont{Pollachini et~al.}(2021)\citenamefont{Pollachini,
  Salazar, G{\'o}es, Maciel, and Duzzioni}}]{pollachini2021hybrid}
\bibinfo{author}{\bibfnamefont{G.~G.} \bibnamefont{Pollachini}},
  \bibinfo{author}{\bibfnamefont{J.~P.} \bibnamefont{Salazar}},
  \bibinfo{author}{\bibfnamefont{C.~B.} \bibnamefont{G{\'o}es}},
  \bibinfo{author}{\bibfnamefont{T.~O.} \bibnamefont{Maciel}},
  \bibnamefont{and} \bibinfo{author}{\bibfnamefont{E.~I.}
  \bibnamefont{Duzzioni}}, \bibinfo{journal}{Phys. Rev. A}
  \textbf{\bibinfo{volume}{104}}, \bibinfo{pages}{032426}
  (\bibinfo{year}{2021}).

\bibitem[{\citenamefont{Perona and Malik}(1990)}]{perona1990scale}
\bibinfo{author}{\bibfnamefont{P.}~\bibnamefont{Perona}} \bibnamefont{and}
  \bibinfo{author}{\bibfnamefont{J.}~\bibnamefont{Malik}},
  \bibinfo{journal}{TPAMI} \textbf{\bibinfo{volume}{12}}, \bibinfo{pages}{629}
  (\bibinfo{year}{1990}).

\bibitem[{\citenamefont{Carslaw and Jaeger}(1959)}]{carslaw1959conduction}
\bibinfo{author}{\bibfnamefont{H.~S.} \bibnamefont{Carslaw}} \bibnamefont{and}
  \bibinfo{author}{\bibfnamefont{J.}~\bibnamefont{Jaeger}},
  \bibinfo{journal}{New York} \textbf{\bibinfo{volume}{510}}
  (\bibinfo{year}{1959}).

\bibitem[{\citenamefont{Wilmott et~al.}(1995)\citenamefont{Wilmott, Howison,
  and Dewynne}}]{wilmott1995mathematics}
\bibinfo{author}{\bibfnamefont{P.}~\bibnamefont{Wilmott}},
  \bibinfo{author}{\bibfnamefont{S.}~\bibnamefont{Howison}}, \bibnamefont{and}
  \bibinfo{author}{\bibfnamefont{J.}~\bibnamefont{Dewynne}},
  \emph{\bibinfo{title}{The mathematics of financial derivatives: a student
  introduction}} (\bibinfo{publisher}{Cambridge university press},
  \bibinfo{year}{1995}).

\bibitem[{\citenamefont{Harrow et~al.}(2009{\natexlab{a}})\citenamefont{Harrow,
  Hassidim, and Lloyd}}]{PhysRevLett.103.150502}
\bibinfo{author}{\bibfnamefont{A.~W.} \bibnamefont{Harrow}},
  \bibinfo{author}{\bibfnamefont{A.}~\bibnamefont{Hassidim}}, \bibnamefont{and}
  \bibinfo{author}{\bibfnamefont{S.}~\bibnamefont{Lloyd}},
  \bibinfo{journal}{Phys. Rev. Lett.} \textbf{\bibinfo{volume}{103}},
  \bibinfo{pages}{150502} (\bibinfo{year}{2009}{\natexlab{a}}).

\bibitem[{\citenamefont{Duan et~al.}(2020)\citenamefont{Duan, Yuan, Yu, Huang,
  and Hsieh}}]{duan2020survey}
\bibinfo{author}{\bibfnamefont{B.}~\bibnamefont{Duan}},
  \bibinfo{author}{\bibfnamefont{J.}~\bibnamefont{Yuan}},
  \bibinfo{author}{\bibfnamefont{C.-H.} \bibnamefont{Yu}},
  \bibinfo{author}{\bibfnamefont{J.}~\bibnamefont{Huang}}, \bibnamefont{and}
  \bibinfo{author}{\bibfnamefont{C.-Y.} \bibnamefont{Hsieh}},
  \bibinfo{journal}{Phys. Lett. A} \textbf{\bibinfo{volume}{384}},
  \bibinfo{pages}{126595} (\bibinfo{year}{2020}).

\bibitem[{\citenamefont{Wang et~al.}(2020)\citenamefont{Wang, Wang, Li, Fan,
  Wei, and Gu}}]{wang2020quantum}
\bibinfo{author}{\bibfnamefont{S.}~\bibnamefont{Wang}},
  \bibinfo{author}{\bibfnamefont{Z.}~\bibnamefont{Wang}},
  \bibinfo{author}{\bibfnamefont{W.}~\bibnamefont{Li}},
  \bibinfo{author}{\bibfnamefont{L.}~\bibnamefont{Fan}},
  \bibinfo{author}{\bibfnamefont{Z.}~\bibnamefont{Wei}}, \bibnamefont{and}
  \bibinfo{author}{\bibfnamefont{Y.}~\bibnamefont{Gu}},
  \bibinfo{journal}{Quant. Inf. Proc.} \textbf{\bibinfo{volume}{19}},
  \bibinfo{pages}{1} (\bibinfo{year}{2020}).

\bibitem[{\citenamefont{Montanaro and Pallister}(2016)}]{montanaro2016quantum}
\bibinfo{author}{\bibfnamefont{A.}~\bibnamefont{Montanaro}} \bibnamefont{and}
  \bibinfo{author}{\bibfnamefont{S.}~\bibnamefont{Pallister}},
  \bibinfo{journal}{Phys. Rev. A} \textbf{\bibinfo{volume}{93}},
  \bibinfo{pages}{032324} (\bibinfo{year}{2016}).

\bibitem[{\citenamefont{Clader et~al.}(2013)\citenamefont{Clader, Jacobs, and
  Sprouse}}]{clader2013preconditioned}
\bibinfo{author}{\bibfnamefont{B.~D.} \bibnamefont{Clader}},
  \bibinfo{author}{\bibfnamefont{B.~C.} \bibnamefont{Jacobs}},
  \bibnamefont{and} \bibinfo{author}{\bibfnamefont{C.~R.}
  \bibnamefont{Sprouse}}, \bibinfo{journal}{Phys. Rev. Lett.}
  \textbf{\bibinfo{volume}{110}}, \bibinfo{pages}{250504}
  (\bibinfo{year}{2013}).

\bibitem[{\citenamefont{Bravo-Prieto et~al.}(2019)\citenamefont{Bravo-Prieto,
  LaRose, Cerezo, Subasi, Cincio, and Coles}}]{bravo2019variational}
\bibinfo{author}{\bibfnamefont{C.}~\bibnamefont{Bravo-Prieto}},
  \bibinfo{author}{\bibfnamefont{R.}~\bibnamefont{LaRose}},
  \bibinfo{author}{\bibfnamefont{M.}~\bibnamefont{Cerezo}},
  \bibinfo{author}{\bibfnamefont{Y.}~\bibnamefont{Subasi}},
  \bibinfo{author}{\bibfnamefont{L.}~\bibnamefont{Cincio}}, \bibnamefont{and}
  \bibinfo{author}{\bibfnamefont{P.~J.} \bibnamefont{Coles}},
  \bibinfo{journal}{arXiv preprint arXiv:1909.05820}  (\bibinfo{year}{2019}).

\bibitem[{\citenamefont{Cerezo et~al.}(2021)\citenamefont{Cerezo, Arrasmith,
  Babbush, Benjamin, Endo, Fujii, McClean, Mitarai, Yuan, Cincio
  et~al.}}]{cerezo2021variational}
\bibinfo{author}{\bibfnamefont{M.}~\bibnamefont{Cerezo}},
  \bibinfo{author}{\bibfnamefont{A.}~\bibnamefont{Arrasmith}},
  \bibinfo{author}{\bibfnamefont{R.}~\bibnamefont{Babbush}},
  \bibinfo{author}{\bibfnamefont{S.~C.} \bibnamefont{Benjamin}},
  \bibinfo{author}{\bibfnamefont{S.}~\bibnamefont{Endo}},
  \bibinfo{author}{\bibfnamefont{K.}~\bibnamefont{Fujii}},
  \bibinfo{author}{\bibfnamefont{J.~R.} \bibnamefont{McClean}},
  \bibinfo{author}{\bibfnamefont{K.}~\bibnamefont{Mitarai}},
  \bibinfo{author}{\bibfnamefont{X.}~\bibnamefont{Yuan}},
  \bibinfo{author}{\bibfnamefont{L.}~\bibnamefont{Cincio}},
  \bibnamefont{et~al.}, \bibinfo{journal}{Nature Reviews Physics}
  \textbf{\bibinfo{volume}{3}}, \bibinfo{pages}{625} (\bibinfo{year}{2021}).

\bibitem[{\citenamefont{Fontanela et~al.}(2021)\citenamefont{Fontanela,
  Jacquier, and Oumgari}}]{fontanela2021quantum}
\bibinfo{author}{\bibfnamefont{F.}~\bibnamefont{Fontanela}},
  \bibinfo{author}{\bibfnamefont{A.}~\bibnamefont{Jacquier}}, \bibnamefont{and}
  \bibinfo{author}{\bibfnamefont{M.}~\bibnamefont{Oumgari}},
  \bibinfo{journal}{SIAM J. Financ. Math.} \textbf{\bibinfo{volume}{12}},
  \bibinfo{pages}{SC98} (\bibinfo{year}{2021}).

\bibitem[{\citenamefont{Fontana et~al.}(2021)\citenamefont{Fontana,
  Fitzpatrick, Ramo, Duncan, and Rungger}}]{fontana2021evaluating}
\bibinfo{author}{\bibfnamefont{E.}~\bibnamefont{Fontana}},
  \bibinfo{author}{\bibfnamefont{N.}~\bibnamefont{Fitzpatrick}},
  \bibinfo{author}{\bibfnamefont{D.~M.} \bibnamefont{Ramo}},
  \bibinfo{author}{\bibfnamefont{R.}~\bibnamefont{Duncan}}, \bibnamefont{and}
  \bibinfo{author}{\bibfnamefont{I.}~\bibnamefont{Rungger}},
  \bibinfo{journal}{Phys. Rev. A} \textbf{\bibinfo{volume}{104}},
  \bibinfo{pages}{022403} (\bibinfo{year}{2021}).

\bibitem[{\citenamefont{Kubo et~al.}(2021)\citenamefont{Kubo, Nakagawa, Endo,
  and Nagayama}}]{kubo2021variational}
\bibinfo{author}{\bibfnamefont{K.}~\bibnamefont{Kubo}},
  \bibinfo{author}{\bibfnamefont{Y.~O.} \bibnamefont{Nakagawa}},
  \bibinfo{author}{\bibfnamefont{S.}~\bibnamefont{Endo}}, \bibnamefont{and}
  \bibinfo{author}{\bibfnamefont{S.}~\bibnamefont{Nagayama}},
  \bibinfo{journal}{Phys. Rev. A} \textbf{\bibinfo{volume}{103}},
  \bibinfo{pages}{052425} (\bibinfo{year}{2021}).

\bibitem[{\citenamefont{Lubasch et~al.}(2020)\citenamefont{Lubasch, Joo,
  Moinier, Kiffner, and Jaksch}}]{lubasch2020variational}
\bibinfo{author}{\bibfnamefont{M.}~\bibnamefont{Lubasch}},
  \bibinfo{author}{\bibfnamefont{J.}~\bibnamefont{Joo}},
  \bibinfo{author}{\bibfnamefont{P.}~\bibnamefont{Moinier}},
  \bibinfo{author}{\bibfnamefont{M.}~\bibnamefont{Kiffner}}, \bibnamefont{and}
  \bibinfo{author}{\bibfnamefont{D.}~\bibnamefont{Jaksch}},
  \bibinfo{journal}{Phys. Rev. A} \textbf{\bibinfo{volume}{101}},
  \bibinfo{pages}{010301(R)} (\bibinfo{year}{2020}).

\bibitem[{\citenamefont{Yang et~al.}(2021)\citenamefont{Yang, Shan, Zhao, and
  Xu}}]{yang2021variational}
\bibinfo{author}{\bibfnamefont{Y.}~\bibnamefont{Yang}},
  \bibinfo{author}{\bibfnamefont{Z.}~\bibnamefont{Shan}},
  \bibinfo{author}{\bibfnamefont{B.}~\bibnamefont{Zhao}}, \bibnamefont{and}
  \bibinfo{author}{\bibfnamefont{L.}~\bibnamefont{Xu}}, \bibinfo{journal}{J.
  Phys.: Conf. Ser.} \textbf{\bibinfo{volume}{1883}}, \bibinfo{pages}{012007}
  (\bibinfo{year}{2021}).

\bibitem[{\citenamefont{Preskill}(2018)}]{Preskill2018quantumcomputingin}
\bibinfo{author}{\bibfnamefont{J.}~\bibnamefont{Preskill}},
  \bibinfo{journal}{{Quantum}} \textbf{\bibinfo{volume}{2}},
  \bibinfo{pages}{79} (\bibinfo{year}{2018}).

\bibitem[{\citenamefont{Wang et~al.}(2021)\citenamefont{Wang, Fontana, Cerezo,
  Sharma, Sone, Cincio, and Coles}}]{wang2021noise}
\bibinfo{author}{\bibfnamefont{S.}~\bibnamefont{Wang}},
  \bibinfo{author}{\bibfnamefont{E.}~\bibnamefont{Fontana}},
  \bibinfo{author}{\bibfnamefont{M.}~\bibnamefont{Cerezo}},
  \bibinfo{author}{\bibfnamefont{K.}~\bibnamefont{Sharma}},
  \bibinfo{author}{\bibfnamefont{A.}~\bibnamefont{Sone}},
  \bibinfo{author}{\bibfnamefont{L.}~\bibnamefont{Cincio}}, \bibnamefont{and}
  \bibinfo{author}{\bibfnamefont{P.~J.} \bibnamefont{Coles}},
  \bibinfo{journal}{Nat. Com.} \textbf{\bibinfo{volume}{12}},
  \bibinfo{pages}{1} (\bibinfo{year}{2021}).

\bibitem[{\citenamefont{Holmes et~al.}(2022)\citenamefont{Holmes, Sharma,
  Cerezo, and Coles}}]{holmes2022connecting}
\bibinfo{author}{\bibfnamefont{Z.}~\bibnamefont{Holmes}},
  \bibinfo{author}{\bibfnamefont{K.}~\bibnamefont{Sharma}},
  \bibinfo{author}{\bibfnamefont{M.}~\bibnamefont{Cerezo}}, \bibnamefont{and}
  \bibinfo{author}{\bibfnamefont{P.~J.} \bibnamefont{Coles}},
  \bibinfo{journal}{PRX Quantum} \textbf{\bibinfo{volume}{3}},
  \bibinfo{pages}{010313} (\bibinfo{year}{2022}).

\bibitem[{\citenamefont{Devitt et~al.}(2013)\citenamefont{Devitt, Munro, and
  Nemoto}}]{devitt2013quantum}
\bibinfo{author}{\bibfnamefont{S.~J.} \bibnamefont{Devitt}},
  \bibinfo{author}{\bibfnamefont{W.~J.} \bibnamefont{Munro}}, \bibnamefont{and}
  \bibinfo{author}{\bibfnamefont{K.}~\bibnamefont{Nemoto}},
  \bibinfo{journal}{Rep. Prog. Phys.} \textbf{\bibinfo{volume}{76}},
  \bibinfo{pages}{076001} (\bibinfo{year}{2013}).

\bibitem[{\citenamefont{Fowler et~al.}(2012)\citenamefont{Fowler, Mariantoni,
  Martinis, and Cleland}}]{fowler2012surface}
\bibinfo{author}{\bibfnamefont{A.~G.} \bibnamefont{Fowler}},
  \bibinfo{author}{\bibfnamefont{M.}~\bibnamefont{Mariantoni}},
  \bibinfo{author}{\bibfnamefont{J.~M.} \bibnamefont{Martinis}},
  \bibnamefont{and} \bibinfo{author}{\bibfnamefont{A.~N.}
  \bibnamefont{Cleland}}, \bibinfo{journal}{Phys. Rev. A}
  \textbf{\bibinfo{volume}{86}}, \bibinfo{pages}{032324}
  (\bibinfo{year}{2012}).

\bibitem[{\citenamefont{Pino et~al.}(2021)\citenamefont{Pino, Dreiling,
  Figgatt, Gaebler, Moses, Allman, Baldwin, Foss-Feig, Hayes, Mayer
  et~al.}}]{pino2020demonstration}
\bibinfo{author}{\bibfnamefont{J.~M.} \bibnamefont{Pino}},
  \bibinfo{author}{\bibfnamefont{J.~M.} \bibnamefont{Dreiling}},
  \bibinfo{author}{\bibfnamefont{C.}~\bibnamefont{Figgatt}},
  \bibinfo{author}{\bibfnamefont{J.~P.} \bibnamefont{Gaebler}},
  \bibinfo{author}{\bibfnamefont{S.~A.} \bibnamefont{Moses}},
  \bibinfo{author}{\bibfnamefont{M.}~\bibnamefont{Allman}},
  \bibinfo{author}{\bibfnamefont{C.}~\bibnamefont{Baldwin}},
  \bibinfo{author}{\bibfnamefont{M.}~\bibnamefont{Foss-Feig}},
  \bibinfo{author}{\bibfnamefont{D.}~\bibnamefont{Hayes}},
  \bibinfo{author}{\bibfnamefont{K.}~\bibnamefont{Mayer}},
  \bibnamefont{et~al.}, \bibinfo{journal}{Nature}
  \textbf{\bibinfo{volume}{592}}, \bibinfo{pages}{209} (\bibinfo{year}{2021}).

\bibitem[{\citenamefont{Grzesiak et~al.}(2020)\citenamefont{Grzesiak,
  Bl{\"u}mel, Wright, Beck, Pisenti, Li, Chaplin, Amini, Debnath, Chen
  et~al.}}]{grzesiak2020efficient}
\bibinfo{author}{\bibfnamefont{N.}~\bibnamefont{Grzesiak}},
  \bibinfo{author}{\bibfnamefont{R.}~\bibnamefont{Bl{\"u}mel}},
  \bibinfo{author}{\bibfnamefont{K.}~\bibnamefont{Wright}},
  \bibinfo{author}{\bibfnamefont{K.~M.} \bibnamefont{Beck}},
  \bibinfo{author}{\bibfnamefont{N.~C.} \bibnamefont{Pisenti}},
  \bibinfo{author}{\bibfnamefont{M.}~\bibnamefont{Li}},
  \bibinfo{author}{\bibfnamefont{V.}~\bibnamefont{Chaplin}},
  \bibinfo{author}{\bibfnamefont{J.~M.} \bibnamefont{Amini}},
  \bibinfo{author}{\bibfnamefont{S.}~\bibnamefont{Debnath}},
  \bibinfo{author}{\bibfnamefont{J.-S.} \bibnamefont{Chen}},
  \bibnamefont{et~al.}, \bibinfo{journal}{Nat Commun}
  \textbf{\bibinfo{volume}{11}}, \bibinfo{pages}{2963} (\bibinfo{year}{2020}).

\bibitem[{\citenamefont{Bravyi et~al.}(2022)\citenamefont{Bravyi, Dial,
  Gambetta, Gil, and Nazario}}]{bravyi2022future}
\bibinfo{author}{\bibfnamefont{S.}~\bibnamefont{Bravyi}},
  \bibinfo{author}{\bibfnamefont{O.}~\bibnamefont{Dial}},
  \bibinfo{author}{\bibfnamefont{J.~M.} \bibnamefont{Gambetta}},
  \bibinfo{author}{\bibfnamefont{D.}~\bibnamefont{Gil}}, \bibnamefont{and}
  \bibinfo{author}{\bibfnamefont{Z.}~\bibnamefont{Nazario}},
  \bibinfo{journal}{J Appl Phys} \textbf{\bibinfo{volume}{132}}
  (\bibinfo{year}{2022}).

\bibitem[{\citenamefont{Gambetta}(2020)}]{gambetta2020ibm}
\bibinfo{author}{\bibfnamefont{J.}~\bibnamefont{Gambetta}},
  \bibinfo{journal}{IBM Research Blog (September 2020)}
  (\bibinfo{year}{2020}).

\bibitem[{\citenamefont{Kim et~al.}(2023)\citenamefont{Kim, Eddins, Anand, Wei,
  Van Den~Berg, Rosenblatt, Nayfeh, Wu, Zaletel, Temme
  et~al.}}]{kim2023evidence}
\bibinfo{author}{\bibfnamefont{Y.}~\bibnamefont{Kim}},
  \bibinfo{author}{\bibfnamefont{A.}~\bibnamefont{Eddins}},
  \bibinfo{author}{\bibfnamefont{S.}~\bibnamefont{Anand}},
  \bibinfo{author}{\bibfnamefont{K.~X.} \bibnamefont{Wei}},
  \bibinfo{author}{\bibfnamefont{E.}~\bibnamefont{Van Den~Berg}},
  \bibinfo{author}{\bibfnamefont{S.}~\bibnamefont{Rosenblatt}},
  \bibinfo{author}{\bibfnamefont{H.}~\bibnamefont{Nayfeh}},
  \bibinfo{author}{\bibfnamefont{Y.}~\bibnamefont{Wu}},
  \bibinfo{author}{\bibfnamefont{M.}~\bibnamefont{Zaletel}},
  \bibinfo{author}{\bibfnamefont{K.}~\bibnamefont{Temme}},
  \bibnamefont{et~al.}, \bibinfo{journal}{Nature}
  \textbf{\bibinfo{volume}{618}}, \bibinfo{pages}{500} (\bibinfo{year}{2023}).

\bibitem[{\citenamefont{Huang et~al.}(2021)\citenamefont{Huang, Bharti, and
  Rebentrost}}]{huang2021near}
\bibinfo{author}{\bibfnamefont{H.-Y.} \bibnamefont{Huang}},
  \bibinfo{author}{\bibfnamefont{K.}~\bibnamefont{Bharti}}, \bibnamefont{and}
  \bibinfo{author}{\bibfnamefont{P.}~\bibnamefont{Rebentrost}},
  \bibinfo{journal}{New J. of Phys.} \textbf{\bibinfo{volume}{23}},
  \bibinfo{pages}{113021} (\bibinfo{year}{2021}).

\bibitem[{\citenamefont{Zienkiewicz et~al.}(2005)\citenamefont{Zienkiewicz,
  Taylor, and Zhu}}]{zienkiewicz2005finite}
\bibinfo{author}{\bibfnamefont{O.~C.} \bibnamefont{Zienkiewicz}},
  \bibinfo{author}{\bibfnamefont{R.~L.} \bibnamefont{Taylor}},
  \bibnamefont{and} \bibinfo{author}{\bibfnamefont{J.~Z.} \bibnamefont{Zhu}},
  \emph{\bibinfo{title}{The finite element method: its basis and fundamentals}}
  (\bibinfo{publisher}{Elsevier}, \bibinfo{year}{2005}).

\bibitem[{\citenamefont{{\"O}zi{\c{s}}ik
  et~al.}(2017)\citenamefont{{\"O}zi{\c{s}}ik, Orlande, Cola{\c{c}}o, and
  Cotta}}]{ozicsik2017finite}
\bibinfo{author}{\bibfnamefont{M.~N.} \bibnamefont{{\"O}zi{\c{s}}ik}},
  \bibinfo{author}{\bibfnamefont{H.~R.} \bibnamefont{Orlande}},
  \bibinfo{author}{\bibfnamefont{M.~J.} \bibnamefont{Cola{\c{c}}o}},
  \bibnamefont{and} \bibinfo{author}{\bibfnamefont{R.~M.} \bibnamefont{Cotta}},
  \emph{\bibinfo{title}{Finite difference methods in heat transfer}}
  (\bibinfo{publisher}{CRC press}, \bibinfo{year}{2017}).

\bibitem[{\citenamefont{Higham}(2002)}]{higham2002accuracy}
\bibinfo{author}{\bibfnamefont{N.~J.} \bibnamefont{Higham}},
  \emph{\bibinfo{title}{Accuracy and stability of numerical algorithms}}
  (\bibinfo{publisher}{SIAM}, \bibinfo{year}{2002}).

\bibitem[{\citenamefont{Press}(2007)}]{press2007numerical}
\bibinfo{author}{\bibfnamefont{W.~H.} \bibnamefont{Press}},
  \emph{\bibinfo{title}{Numerical recipes 3rd edition: The art of scientific
  computing}} (\bibinfo{publisher}{Cambridge university press},
  \bibinfo{year}{2007}).

\bibitem[{\citenamefont{Belsley et~al.}(2005)\citenamefont{Belsley, Kuh, and
  Welsch}}]{belsley2005regression}
\bibinfo{author}{\bibfnamefont{D.~A.} \bibnamefont{Belsley}},
  \bibinfo{author}{\bibfnamefont{E.}~\bibnamefont{Kuh}}, \bibnamefont{and}
  \bibinfo{author}{\bibfnamefont{R.~E.} \bibnamefont{Welsch}},
  \emph{\bibinfo{title}{Regression diagnostics: Identifying influential data
  and sources of collinearity}} (\bibinfo{publisher}{John Wiley \& Sons},
  \bibinfo{year}{2005}).

\bibitem[{\citenamefont{Harrow et~al.}(2009{\natexlab{b}})\citenamefont{Harrow,
  Hassidim, and Lloyd}}]{hhl}
\bibinfo{author}{\bibfnamefont{A.~W.} \bibnamefont{Harrow}},
  \bibinfo{author}{\bibfnamefont{A.}~\bibnamefont{Hassidim}}, \bibnamefont{and}
  \bibinfo{author}{\bibfnamefont{S.}~\bibnamefont{Lloyd}},
  \bibinfo{journal}{Phys. Rev. Lett.} \textbf{\bibinfo{volume}{103}},
  \bibinfo{pages}{150502} (\bibinfo{year}{2009}{\natexlab{b}}).

\bibitem[{\citenamefont{Grover and Rudolph}(2002)}]{grover2002creating}
\bibinfo{author}{\bibfnamefont{L.}~\bibnamefont{Grover}} \bibnamefont{and}
  \bibinfo{author}{\bibfnamefont{T.}~\bibnamefont{Rudolph}},
  \bibinfo{journal}{arXiv preprint quant-ph/0208112}  (\bibinfo{year}{2002}).

\bibitem[{\citenamefont{Saha et~al.}(2022)\citenamefont{Saha, Robson,
  Howington, Suh, Wang, and Nabrzyski}}]{saha2022advancing}
\bibinfo{author}{\bibfnamefont{K.~K.} \bibnamefont{Saha}},
  \bibinfo{author}{\bibfnamefont{W.}~\bibnamefont{Robson}},
  \bibinfo{author}{\bibfnamefont{C.}~\bibnamefont{Howington}},
  \bibinfo{author}{\bibfnamefont{I.-S.} \bibnamefont{Suh}},
  \bibinfo{author}{\bibfnamefont{Z.}~\bibnamefont{Wang}}, \bibnamefont{and}
  \bibinfo{author}{\bibfnamefont{J.}~\bibnamefont{Nabrzyski}},
  \bibinfo{journal}{arXiv preprint arXiv:2210.16668}  (\bibinfo{year}{2022}).

\end{thebibliography}

\appendix

\section{Generalization of the heat equation to higher dimensions}\label{app:high dim}

We have considered the case of the heat equation with single spatial dimension. The generalization to the multidimensional case is straightforward. Reference \cite{saha2022advancing} shows how the multidimensional analog of the matrix for Poisson equation scales with the growth of the number of spacial dimensions. We use a similar approach for the multidimensional matrix $A^{(d_r)}(c)$
\begin{eqnarray}
\begin{gathered}
A^{\left(d_r\right)}\left(c\right)=\underbrace{A\left(0\right)\otimes I\otimes I\otimes\dots\otimes I}_{d_r}\\+I\otimes A\left(0\right)\otimes I\otimes\dots\otimes I+\dots\\+I\otimes I\otimes I\otimes\dots\otimes A\left(0\right)-cI^{\left(2^nd_r\right)},   
\end{gathered}
\label{A_coordinate_scaling}
\end{eqnarray}
where $I$ is an identity gate for $n$ qubits. This form of the matrix $A^{(d_r)}$ allows us to determine the effective set of $U_i$ for ATA described in the Section \ref{s:prelim}. For example, the $i$th term of the sum (\ref{A_coordinate_scaling}) can be decomposed as
\begin{eqnarray}
\begin{gathered}
\underbrace{I\otimes I\otimes I\otimes\dots\otimes I}_{i-1}\otimes A\otimes\dots\otimes I\\
=\underbrace{I\otimes I\otimes I\otimes\dots\otimes I}_{i-1}\otimes \sum_{i=1}^{K_A}\beta_iU_i\otimes\dots\otimes I,
\label{i_A_coordinate_scaling}
\end{gathered}
\end{eqnarray}
where the decomposition of the matrix A (\ref{A_ansatz_tree_decomposition}) has been used. Hence, ATA scales as {$\mathcal{O}(d_r^6d^2n^4)$} depending on the dimension of the coordinate space $d_r$, where $n$ is the number of qubits required for a single spatial dimension. In practical terms, this means that to solve the heat equation with, say, three spatial dimensions, we would need to use three times as many qubits, and the depth of the tree as well as the depth of the circuits would grow at a similar rate.

\section{Arbitrary Ansatz with $I$ and $Z$ Pauli matrices}\label{app: arbitrary with I and Z}

In this Appendix we demonstrate that an arbitrary vector could be constructed using decomposition
\begin{eqnarray}
\begin{gathered}
    \psi=M_{\mbox{QFT}}^\dagger
    \left(\sum_{i=0}^{2^n-1}h_i\Pi_{_i}\right)
    M_{\mbox{QFT}}\ket{0}^{\otimes n},\\
\end{gathered}
\label{arbitrary_vec}
\end{eqnarray}
where Pauli products $\Pi_i$ consist only of the tensor product of $I$ and $Z$ for $n$ qubits, $Z$ gates apply where $1$ takes place in the binary decomposition of $i$. First, we consider the inner part of the decomposition
\begin{eqnarray}
M_D=\sum_{p=0}^{2^n-1}h_p\Pi_{_p}.
\label{inner_part}
\end{eqnarray}
Note that $M_D$ is a diagonal matrix, since a tensor product of $I$ and $Z$ is always a diagonal matrix. In addition, we note that the form (\ref{inner_part}) can create any diagonal matrix using the formula for $h_p$
\begin{eqnarray}
h_p=\sum_{i=0}^{2^n-1}\lambda^{M_D}_i(-1)^{\sum_si_sp_s},
\label{coeeficients_of_decomposition_to_Pauli_products}
\end{eqnarray}
where $p_s$ and $i_s$ are the $s$th binary digit of $p$ and $i$ index, respectively; $\lambda^{M_D}_i=M_{Dii}$ are the eigenvalues of $M_D$. 

Next we focus on a way of getting an arbitrary vector using 
\begin{eqnarray}
\begin{gathered}
    \psi_F=
    \left(\sum_{i=0}^{2^n-1}h_i\Pi_{_i}\right)
    M_{\mbox{QFT}}\ket{0}^{\otimes n}\\
    =\sum_{i=0}^{2^n-1}h_i\Pi_{_i}\ket{+}^{\otimes n}=M_D\ket{+}^{\otimes n}=\vec{\lambda}^{M_D}/2^{n-1}.\\
\end{gathered}
\label{arbitrary_vec_fourier}
\end{eqnarray}
Thus, Eq.~(\ref{arbitrary_vec_fourier}) produces a vector that coincides with the diagonal elements of $M_D$. Similarly, Eq.~(\ref{arbitrary_vec}) creates an arbitrary vector, since $M_{\mbox{QFT}^\dagger}$ is a unitary operator that changes the basis. Also note that $\psi_F$ is a Fourier image of $\psi$.

\end{document}